# Benchmarking Physical Performance of Neural Inference Circuits

Dmitri E. Nikonov and Ian A. Young
Components Research, Intel Corp., Hillsboro, Oregon 97007, USA
dmitri.e.nikonov@intel.com

Abstract

Numerous neural network circuits and architectures are presently under active research for application to artificial intelligence and machine learning. Their physical performance metrics (area, time, energy) are estimated. Various types of neural networks (artificial, cellular, spiking, and oscillator) are implemented with multiple CMOS and beyond-CMOS (spintronic, ferroelectric, resistive memory) devices. A consistent and transparent methodology is proposed and used to benchmark this comprehensive set of options across several application cases. Promising architecture/device combinations are identified.

## 1. Introduction

The unprecedented progress of traditional, Boolean computing over the last five decades has been propelled by the scaling of the transistor scaling according to Moore's law [1]. Recently a larger share of computing is being consumed by applications related to artificial intelligence (AI) and machine learning (ML). For these, Boolean computing is less efficient. This has spurred research in neural computing which covers a wide field of research; from neural network algorithms which can be programmed on traditional Boolean hardware like CPUs or GPUs to neural network circuits implemented in specialized hardware – application specific engines. The former approach presently handles the majority of user needs from the data center to the edge. The latter approach resulted in development and research thrusts in e.g. digital neural accelerators such as [2] (see a review [3]) and neuromorphic (biologically inspired) chips such as [4]. The operation of neuromorphic chips can span a range of circuit implementations from mostly digital to mostly analog (see reviews [5] and [6]).

In the last few years, AI/ML achieved prominent successes, especially related to deep neural networks (DNN) [7] and convolutional neural networks (CoNN). ML has enabled a revolutionary improvement in the accuracy of image, pattern, and facial recognition, including the treatment of 'big data' online. More demand for neural computing is emerging in robotic control, autonomous vehicles, drones, etc.

One of the main concerns on the minds of developers of AI computing systems is the same as for traditional computing: the power consumption in the chips. The history of traditional computing shows that the commercial success of computing devices and architectures is predicated largely on their physical performance – areal density, speed of operation, and consumed energy as benchmarked in [8,9]. These ultimately translate into processing throughput and consumed power of the chips, which are of utmost importance to the user. A fair comparison between the published neural network implementations is difficult due to the difference in the process technology generation, the network architectures, and computing workloads.

The main purpose of the paper is to establish a methodology for comparing various neural network hardware approaches and to understand the trends revealed through its development. In doing that we strive to adhere to the following principles:

a) general: wide scope of technologies, devices and circuits;
b) transparent: simple analytics more important than precise simulations;
c) uniform: consistent inputs and assumptions across multiple types of hardware;
d) transparent: all models used are described and the code is available [10] to the reader for verification.

Let us differentiate this work from the existing body of literature. We do not aim to give a literature review, and refer the reader to the excellent review papers in the neuromorphic hardware field [11,12] which do not attempt to quantitatively compare prior works, as we do in this paper. Oftentimes benchmarking refers to comparing various algorithms for an application mainly based on their accuracy and with little reference to hardware implementation, e.g. [13]. In contrast, we compare different types of hardware implementing the same algorithm and focusing on its energy consumption and performance.

The discussion of neural networks in the numerous papers currently being published remains at the architecture level. For example, the accuracy of recognition is studied in its dependence on the number of network elements, topology, and details of the algorithm. We are cognizant of the importance of the accuracy of inferencing. And indeed prior studies discovered that this accuracy can be degraded compared to the algorithm-limited "maximum accuracy" due to device non-idealities [14]. However for this paper we focus on the effect of the different types of devices and their neural circuits. For that purpose we make an optimistic assumption of devices not degrading accuracy, as exemplified by [15]. This assumption is appropriate for benchmarking which targets the idealized, paradigm limiting cases.

Some research papers report experimentally measured performance and energy in several implementations of CoNN [16,17] by running them on a GPU or a particular DNN on a variety of neural hardware [18]. In contrast, we provide a theoretical prediction approach to benchmarking. A rigorous simulation framework, NeuroSim, is used to benchmark the neural network circuit architecture in a cross-connect topology based on a set of memory cells serving as synapses [19]. The Eyeriss simulation tool focuses on accelerators for DNN [20]. The CrossSim simulator has similar capabilities [21].

Benchmarking of a variety of devices, including beyond-CMOS ones, has been done in an approach similar to ours, but for one type of neural networks only, cellular neural networks (CeNN) [22]. Various estimates of time and energy of operation for certain types of digital and analog devices have been done previously [23,24,25,26]. Compared to these prior works, w cover a wider scope of devices and network architectures with less detailed models of circuits.

Another purpose of this paper is to explore the impact of exploratory devices on the performance and energy efficiency of operation for neural networks. For example, neural circuits based on the spintronic type of beyond CMOS devices have been proposed [27,28]. In this paper we aim to expand the list of beyond CMOS devices applied to neural networks. Also we consider both digital and analog neural networks in an attempt to understand whether there is an advantage in speed and energy of neural computing with the latter. We also consider several types of neural network microarchitectures and analyze their relative advantages. Finally we consider several cases of neural networks running their application 'workloads' and demonstrate that the qualitative conclusions made about various neural network hardware remains valid when performing with these "real use" cases.

## 2. Fundamentals and Concepts of Neuromorphic Computing

Operation in the majority of neural network architectures relies on a neural gate, often called the perceptron, Figure 1. The elements at the input, synapses, receive vectors of input signals, $x_i$, and multiply them by vectors of weights, $w_i$. Neurons perform the summation of these products and apply a nonlinear threshold (or 'activation') function.

$$f(x) = g\left(\sum_i^n w_i x_i + b\right) \quad (1)$$

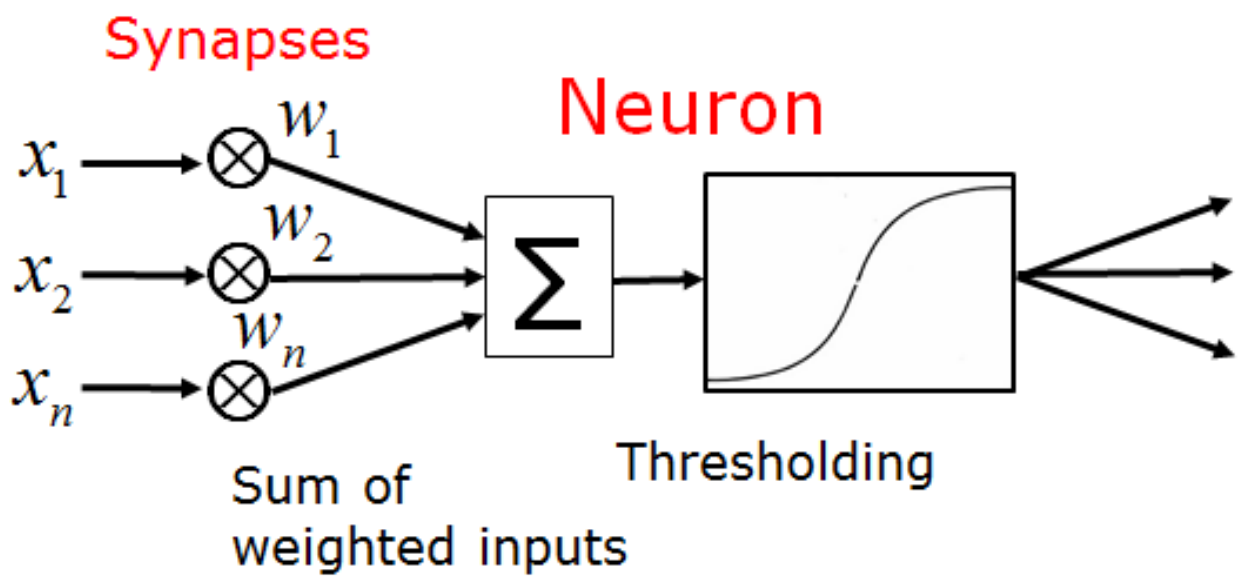

Figure 1. Scheme of a neural gate, perceptron.

Despite its apparent simplicity, the neural gate in some form underlies most of the neuromorphic hardware and algorithms. Deep neural networks (DNN) consist of cascaded multiple layers of neural gates. Convolutional neural networks (CoNN) are an example of algorithms applying DNN to image processing. For the benchmarking analysis of this paper we only consider the DNN workloads of inference (i.e., determining a distance of input vectors from memorized ones in a multi-dimensional space). Inference is crucial for recognition, i.e., classifying objects in the input data.

Learning (or, training) is the process of modifying parameters of the neural network for better recognition. It consists of performing numerous inferences on the input data and then adjusting the weights in the neural network. Methods of learning can be e.g.: a) supervised learning: optimization of weights through e.g. backpropagation algorithm [7], or b) unsupervised learning: change of weights according to synapse activity caused by input patterns, e.g. using the spike-timing-dependent plasticity (STDP) algorithm [27].

We decided to limit the scope of the present paper to inferencing. We realize the importance of learning and that it requires much more computing effort. However inference and learning present different market segments and have different usage models: learning is mostly practiced by providers of data center services and inference mostly client users. As such it is possible to consider inference and learning separately. Benchmarking of learning (such as in [19]) will be explored in a future publication.

Table 1. LIST OF NOTATION USED IN THE PAPER.

| Quantity | Symbol | Units | Value |
|---|---|---|---|
| Process generation ('node') size | $F$ | nm | 15 |
| Minimum interconnect length | $l_{ic}$ | nm | $20F$ |
| Bits in a digital synapse | $n_b$ | | 8 |
| Levels in an analog synapse | $n_l$ | | 64 |
| Area, delay, energy of circuits | $a, \tau, E$ | nm$^2$,ps,fJ | |
| Area, delay, energy of a device | $a_{dev}, \tau_{dev}, E_{dev}$ | nm$^2$,ps,fJ | |
| Area, delay, energy of a minimal interconnect | $a_{ic}, \tau_{ic}, E_{ic}$ | nm$^2$,ps,fJ | |
| Area, delay, energy of an inverter | $a_{inv}, \tau_{inv}, E_{inv}$ | nm$^2$,ps,fJ | |
| Area, delay, energy of a 2 input NAND | $a_{nan}, \tau_{nan}, E_{nan}$ | nm$^2$,ps,fJ | |
| Area, delay, energy of a register bit | $a_{reg}, \tau_{reg}, E_{reg}$ | nm$^2$,ps,fJ | |
| Area, delay, energy of a state element | $a_{se}, \tau_{se}, E_{se}$ | nm$^2$,ps,fJ | |
| Area, delay, energy of a 1-bit full adder | $a_1, \tau_1, E_1$ | nm$^2$,ps,fJ | |
| Area, delay, energy of a n-bit full ripple-carry adder | $a_{add}, \tau_{add}, E_{add}$ | nm$^2$,ps,fJ | |
| Area, delay, energy of a synapse | $a_{syn}, \tau_{syn}, E_{syn}$ | nm², ps, fJ | |
| Area, delay, energy of a neuron | $a_{neu}, \tau_{neu}, E_{neu}$ | nm², ps, fJ | |
| Area, delay, energy of a compute workload | $a_{CW}, \tau_{CW}, E_{CW}$ | nm², ps, fJ | |
| Area, delay, energy of the whole chip | $a_{ch}, \tau_{ch}, E_{ch}$ | nm², ps, fJ | |
| Width of a digital transistor | $w_{dt}$ | nm | $4F$ |
| Width of an analog transistor | $w_{at}$ | nm | $16F$ |
| Capacitance of the transistor per unit width | $c_{tran}$ | F/m | |
| On- and off-current in the transistor per unit width | $i_{on}, i_{off}$ | A/m | |
| Saturation voltage of a transistor | $V_{sat}$ | V | 0.3 |
| Linear transconductance of a transistor | $g_{mdt}$ | S | |
| On-state resistance of a transistor | $R_{ondt}$ | $\Omega$ | |
| Supply voltage | $V_{cc}$ | V | 0.8 |
| Capacitance of an interconnect per length | $c_{ic}$ | nF/m | 0.5 |
| Capacitance of a minimum interconnect | $C_{ic}$ | F | $c_{ic}l_{ic}$ |
| Resistance of an interconnect per length | $r_{ic}$ | $G\Omega$ | 2.2 |
| Resistance of a minimum interconnect | $R_{ic}$ | $\Omega$ | 667 |
| Load capacitance for an interconnect | $C_{load}$ | F | |
| Effective resistance of a synapse | $R_{eff}$ | $\Omega$ | |
| Voltage for a sense amplifier | $V_{sa}$ | V | 0.4 |
| Transistor width for a sense amplifier | $w_p, w_n, w_{iso}, w_{en}$ | nm | $\{4, 4, 6.5, 5\}F$ |
| Sense difference and cell read voltages for a voltage sense amplifier | $V_{vsa}, V_{rvsa}$ | V | {0.1,0.5} |
| Analog read circuit row voltage | $V_{row}$ | V | 0.65 |
| Analog read pulse | $\tau_{repu}$ | ns | 1 |
| Width of input, pull-up, output transistors in OTA | $w_{in}, w_{up}, w_{out}$ | nm | $\{10, 5, 10\}F$ |
| Current from a neuron | $I_{neu}$ | A | |
| Factor of extra number of synapses in CeNN | $M_{syncnn}$ | | 4 |
| Factor of extra settling time in CeNN | $M_{stepcnn}$ | | 5 |
| Maximum weight value for CNN | $w_{\max}$ | | 0.23 |
| Average sum of the weights in CeNN cell | $w_{sum}$ | | 1.26 |
| Factor of spike duration | $N_{spi}$ | | 3 |
| Factor of spacing between spikes | $N_{spa}$ | | 3 |
| Number of spikes for a neuron to fire | $N_{fire}$ | | 10 |
| Spiking activity ratio | $r_a$ | | |
| Number of oscillator periods till synchronization | $N_{synch}$ | | 30 |
| Area overhead factor for a synapse | $M_{syn}$ | | 2 |
| Area overhead factor for a neuron | $M_{neu}$ | | 2 |
| Area overhead factor for a core | $M_{cor}$ | | 2 |
| Area overhead factor for the whole chip | $M_{ch}$ | | 2 |
| Cores per chip; value for a nominal chip | $c_{ch}$ | | 64 |
| Neurons per core; value for a nominal chip | $n_{cor}$ | | 256 |
| Number of input neurons per core | $n_{in}$ | | |
| Number of output neurons per core | $n_{out}$ | | |
| Synapses per neuron; value for a nominal chip | $s_{cor}$ | | 256 |
| Synapses per core | $s_{neu}$ | | |
| Number of feature maps in a stage | $f_{st}$ | | |

## 3. Types of Neuromorphic Devices

Synapses and neurons can be implemented by a variety of devices (Table 2):

- Digital CMOS and analog CMOS or tunnel FET (TFET) devices.
- Ferroelectric FET (FEFET) devices.
- Spintronic devices [27] of five types: in-plane and perpendicular spin transfer torque (STT) switches with perpendicular magnetic anisotropy, spin orbit torque (SOT) switches, domain wall (DW) motion device, and magnetoelectric (ME) switched device.
- Resistive memory elements: oxide RRAM, floating gate resistors (flash), phase-change memory, general spintronic and higher-resistance spin-orbit torque resistors (implemented as magnetic tunnel junctions, MTJ), ferroelectric resistors.

Table 2. PARAMETERS FOR DEVICES COMPRISING SYNAPSES AND NEURONS.

| device name | Area, int | Delay, int | Delay, ic | Energy, int | Energy, ic | Ron | Roff |
|---|---|---|---|---|---|---|---|
| units | $nm^2$ | ps | ps | aJ | aJ | kOhm | kOhm |
| CMOSdig | 3600 | 0.50 | 0.36 | 39.29 | 17.73 | | |
| CMOSana | 14400 | **0.50** | **0.21** | 157.16 | 17.73 | | |
| TFETdig | 3600 | 0.79 | 0.66 | 7.86 | 4.43 | | |
| TFETana | 14400 | 0.79 | 0.26 | 31.43 | 4.43 | | |
| FEFET | 14400 | **100.67** | **1.81** | 2319.80 | 17.73 | | |
| STT,pma | 3600 | 763.28 | 501.00 | 96614.00 | 2.49 | | |
| SOT | 7200 | 911.07 | 279.12 | 23918.00 | 1.11 | | |
| DW | 7200 | 528.25 | 93.30 | 7987.10 | 1.11 | | |
| ME | 7200 | 679.91 | 52.09 | 1108.90 | 0.28 | | |
| OxideR | 3600 | 203.85 | 62.17 | 254.81 | 6.92 | 200 | 1000 |
| FloagaR | 7200 | 1019.20 | 310.33 | 1019.20 | 27.70 | 1000 | 100000 |
| PCMR | 3600 | 50.96 | 15.64 | 1019.20 | 27.70 | 50 | 1000 |
| SpinR | 3600 | 3.06 | 1.06 | 91.73 | 2.49 | 3 | 6 |
| SOTR | 7200 | 9.17 | 2.92 | 40.77 | 1.11 | 9 | 30 |
| FER | 4050 | 30.58 | 9.43 | 499.43 | 13.57 | 30 | 30000 |

To determine the area, delay, and energy of devices and circuits, we rely on our benchmarking methodology [8,9] developed for digital circuits. The benchmarks are calculated consistently for multiple devices scaled to the process node size [8], $F = 15nm$. We first determine the values for an 'intrinsic device' (i.e. a transistor, or a nanomagnet, Table 2) and then for simple circuits [9].

For digital technologies we assume that synapses comprise 8-bit registers. For analog technologies we will assume that synapses are accurately set to 64 levels. Understanding that these two are not equivalent in terms of precision, we choose to keep the typical value of 8 for digital precision. We also assume that the accuracy in analog networks is not limited by the number of levels. We take the best case by ignoring non-idealities of the synapse device characteristics. Realistic cases in which device characteristics affect accuracy are considered e.g. in [29].

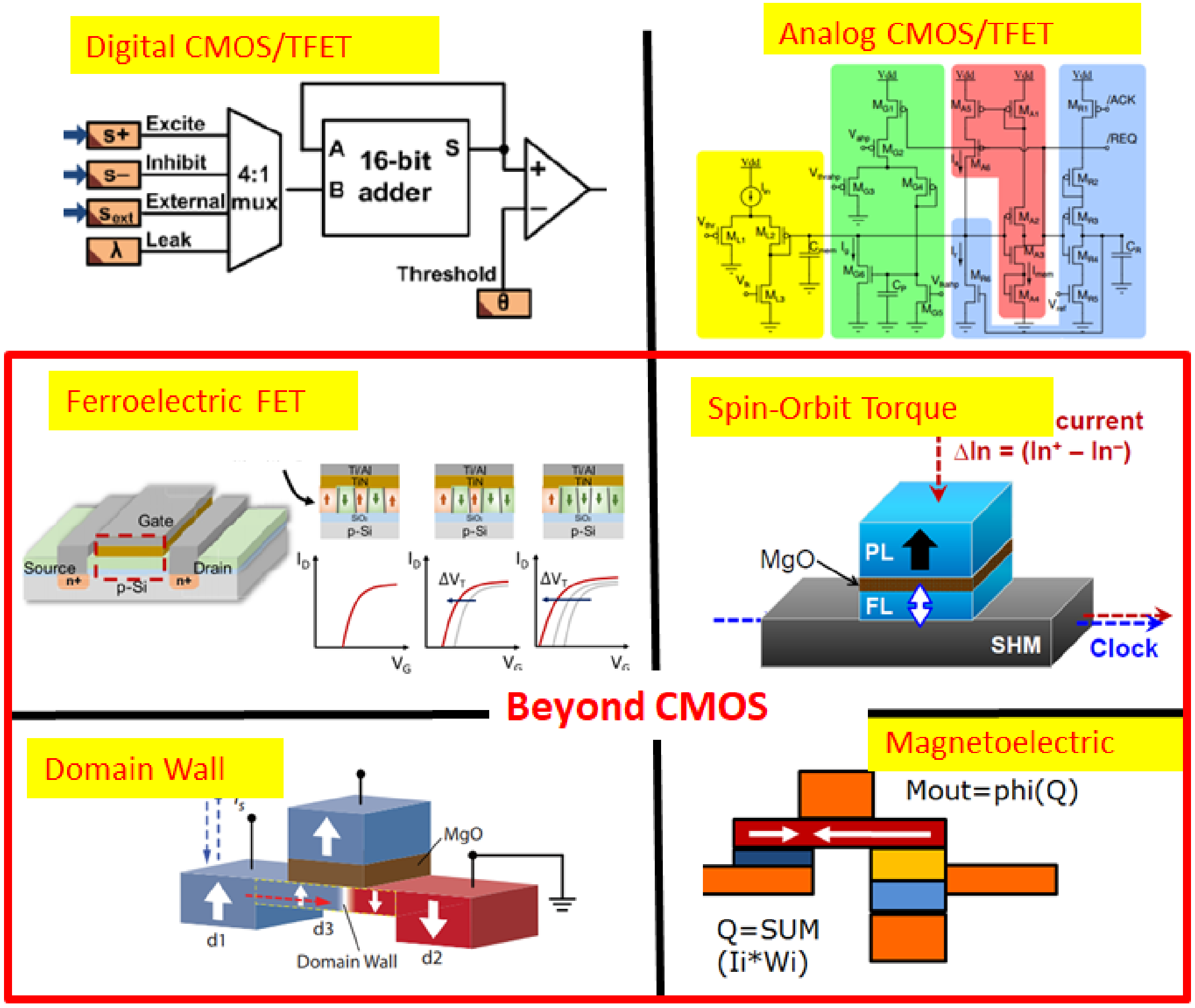

Figure 2. Schemes of neurons implemented with CMOS and beyond-CMOS devices.

Digital CMOS.

The first kind of digital NN is based on SRAM synapses that only provide a weight, while the multiplication and summation (MAC) operations are performed consecutively in the neuron [29]. The circuit considered here follows that in [22]: a synapse consists of n-bits of a SRAM register and state element; a neuron consists of two n-bit registers, an n-bit adder, *n* NAND gates, *n* inverters, and three *n*-state elements. Therefore area of the synapse and the neuron are the sums of the areas of the above constituent circuits. The delay and energy are mostly expended in the neuron, but some of the contributions are proportional to the number of synapses. Therefore such contributions are inserted to the equations for synapses below.

$$a_{syn} = n_b a_{reg} \tag{2}$$

$$\tau_{syn} = 3\tau_{reg} + 4\tau_{se} + \tau_{nan} + \tau_{inv} + n_b\tau_1 \tag{3}$$

$$E_{syn} = n_b\left(3E_{reg} + 4E_{se} + E_{nan} + E_{inv} + E_1\right) \tag{4}$$

$$a_{neu} = n_b\left(2a_{reg} + a_{inv} + a_{nan} + a_1 + a_{se}\right) \tag{5}$$

$$\tau_{neu} = 2\tau_{reg} + 3\tau_{se} + \tau_{nan} + \tau_{inv} + n_b\tau_1 \tag{6}$$

$$E_{neu} = n_b\left(2E_{reg} + 3E_{se} + E_{nan} + E_{inv} + E_1\right) \tag{7}$$

The performance estimate and parameters (Table 1) for a sense amplifier follow [19]. It is used as a part of a reading circuit for SRAM memories. The quantities per bit below are added to the corresponding neuron estimates. The transconductance and load capacitance of

$$a_{sa} = a_{inv1}\left(w_n + w_p + w_{iso} + w_n\right)/\,w_{dt} \tag{8}$$

$$g_{msa} = g_{mdt}\left(w_p + w_n\right)/\,w_{dt} \tag{9}$$

$$C_{lsa} = c_{tran}\left(w_p + w_n\right) \tag{10}$$

$$\tau_{sa} = \log\left(V_{cc}/V_{sa}\right)c_{lsa}/\,g_{msa} + n_b\tau_1 \tag{11}$$

$$E_{sa} = C_{lsa}V_{cc}^2 \tag{12}$$

where the second term in the delay corresponds to the time to enable the sense amp and is proportional to the clock time.

The performance estimate and parameters (Table 1) for a voltage sense amplifier follow [19]. It is used as a part of a reading circuit for digital resistive memories. The quantities per bit below are added to the corresponding neuron estimates. It comprises 3 of n-type and 3 of p-type minimum width transistors.

$$a_{vsa} = 6a_{inv1} \tag{13}$$

the pre-charge resistance, the sense input capacitance, and the bit line capacitance

$$R_{pch} = R_{ondt} \tag{14}$$

$$C_{si} = 2c_{tran} w_{dt} \tag{15}$$

$$C_{li} = s_{neu} c_{ic} l_{ic} \tag{16}$$

$$\tau_{vsa} = 2.3 R_{pch} C_{si} + V_{vsa} \left( C_{si} + C_{li} \right) / \left( V_{rvsa} / R_{on} - V_{rvsa} / R_{off} \right) + 2 n_b \tau_1 \tag{17}$$

$$E_{vsa} = C_{si} V_{cc}^2 \tag{18}$$

<u>Digital MAC.</u>

Another kind of digital NN contains a multiplier and an adder in every synapse, so that the MAC operation is performed in the synapse [66]. The role of neurons is summation of partial results and application of the activation function.

$$a_{syn} = \left( n_b + 1 \right) a_{add} + a_{se} \tag{19}$$

$$\tau_{syn} = \tau_{add} + \tau_{se} \tag{20}$$

$$E_{syn} = \left( n_b + 1 \right) E_{add} / 2 + E_{se} \tag{21}$$

$$a_{neu} = a_{add} + 2 a_{se} + n_b a_{ram} \tag{22}$$

$$\tau_{neu} = \tau_{add} + 2 \tau_{se} + \tau_{ram} \tag{23}$$

$$E_{neu} = E_{add} + 2 E_{se} + n_b E_{ram} \tag{24}$$

The factor $n_b$ in energy and delay would correspond to simple ripple carry adders and multipliers based on them. More efficient designs based adders and multipliers (e.g. carry-save adders) are accounted by an additional factor of 1/2.

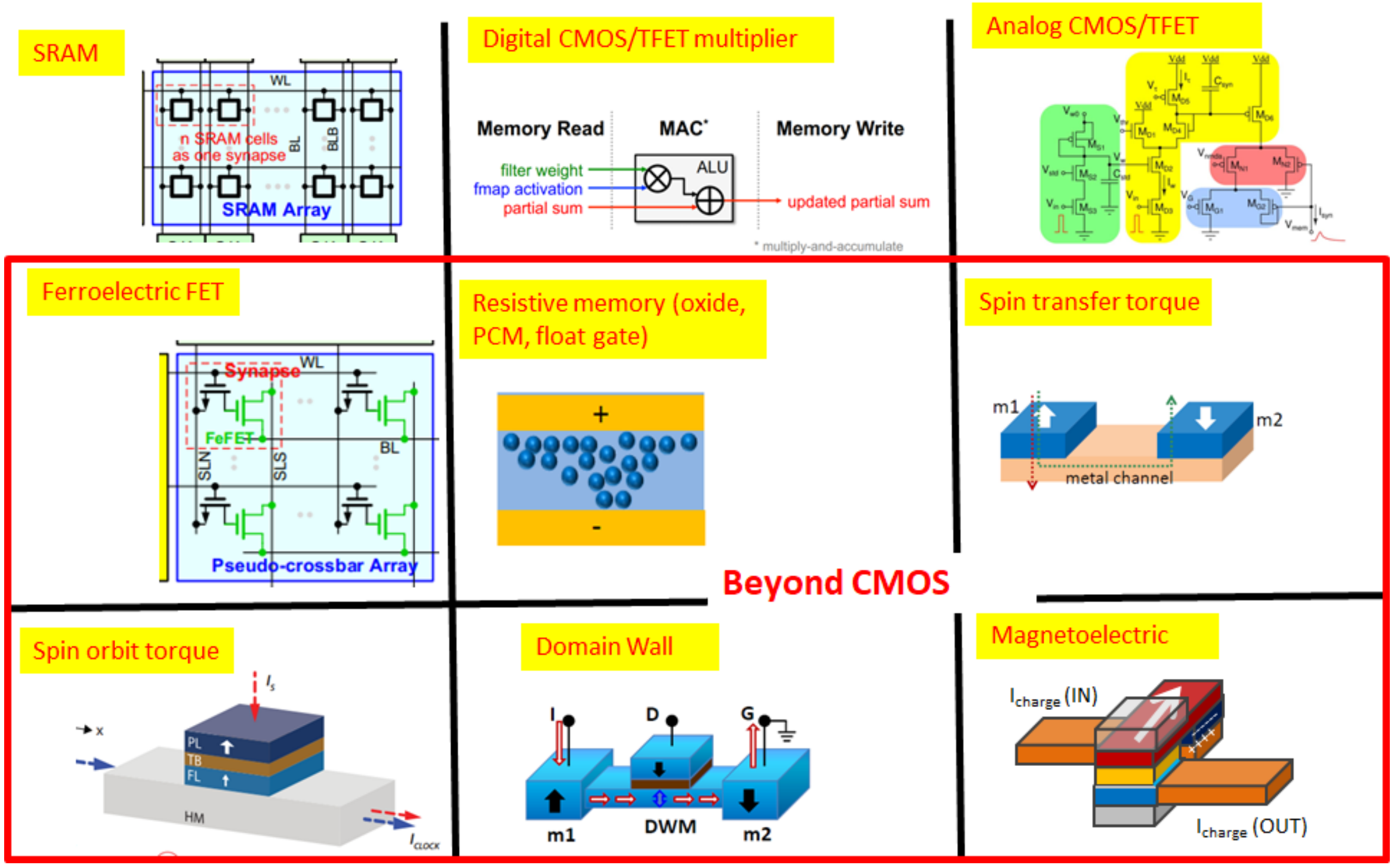

Figure 3. Schemes of synapses implemented with CMOS and beyond-CMOS devices.

## Analog CMOS.

We assume a cell similar to that in [22], where a neuron consists of an opamp, a current source, and a threshold function circuit; a synapse consist of 2 operational transconductance amplifiers (OTA), see also [30]. Transistors of various width are used, Table 1. The effective capacitance of the cell is dominated by the capacitance of the two OTAs

$$C_f = 4c_{tran} w_{out} \tag{25}$$

The subthreshold swing of a transistor is

$$SS = \frac{V_{sat}}{\log_{10}\left(i_{on} / i_{off}\right)}$$

The bias current is approximated as the geometric average of the on- and off-states:

$$I_b = \sqrt{i_{on} i_{off}}\, w_{in} \tag{26}$$

The transconductance of an OTA is

$$g_{mOTA} = \frac{I_b \ln 10}{SS} \frac{w_{out}}{w_{up}} \quad (27)$$

The output conductance of two OTAs is determined by

$$G_m = 2g_{mOTA} / w_{\max} \quad (28)$$

The effective resistance of the cell (with a factor of 2x for the nonlinearity of OTA and 2x to ensure output stability).

$$R_f = 4 / G_m \quad (29)$$

Then the opamp driving current is

$$I_{opamp} = V_{cc} / R_f \quad (30)$$

and the OTA current is

$$I_{OTA} = 2I_b \frac{2w_{sum}}{w_{\max}} \left(1 + \frac{w_{out}}{w_{up}}\right) \quad (31)$$

Thus benchmarks for the synapse and the neuron are

$$a_{syn} = 2a_{inv4}(w_{in} + w_{in} + w_{in}) / w_{idt} \quad (32)$$

$$\tau_{syn} = 8.4 R_f C_f \quad (33)$$

$$P_{syn} = V_{cc} I_{OTA} \quad (34)$$

$$E_{syn} = P_{syn} \tau_{syn} \quad (35)$$

$$a_{syn} = 3a_{inv4}(w_{in} + w_{in} + w_{in}) / w_{idt} \quad (36)$$

$$\tau_{neu} = \tau_{syn} \quad (37)$$

$$P_{neu} = V_{cc}\left(I_{opamp} + I_{OTA}\right) \quad (38)$$

$$E_{neu} = P_{neu} \tau_{neu} \quad (39)$$

where the area of standard cells in circuit is approximated as fan-out-4 inverters.

The performance estimate and parameters (Table 1) for an analog read circuit follow [19]. It is used as a part of a reading circuit for analog valued resistive memories. The quantities per analog cell below are added to the corresponding neuron estimates. It comprises several circuits equal in area to 32 standard inverter cells.

$$a_{adr} = 32a_{inv1} \quad (40)$$

the column voltage is

$$V_{col} = V_{row} - V_{rvsa} \tag{41}$$

$$\tau_{adr} = \tau_{repu} + 2n_b \tau_1 \tag{42}$$

$$P_{adr} = 25V_{col}^2 / R_{ondt} \tag{43}$$

$$E_{adr} = P_{adr} \tau_{adr} \tag{44}$$

<u>Analog spintronic and ferroelectric devices.</u>

Both synapses and neurons consist of just one intrinsic device. Spintronic synapses and neurons have been proposed in [27], as well as ones based on magnetic tunnel junctions [31] or magnetoelectric switching [32]; see overview [33]. We assume the supply voltage to be 0.1V for all spintronic devices. Ferroelectric synapses were explored in [34,35].

These analog neurons and synapses have greater size, delay, and energy proportionally to the number of analog levels:

$$a_{syn} = n_l a_{dev} \tag{45}$$

$$\tau_{syn} = \tau_{dev} \tag{46}$$

$$E_{syn} = E_{dev} \tag{47}$$

$$a_{neu} = n_l a_{dev} \tag{48}$$

$$\tau_{neu} = n_l \tau_{dev} / 4 \tag{49}$$

$$E_{neu} = n_l E_{dev} \tag{50}$$

<u>Resistive memories.</u> We will use this term synonymously with ‘memristor’. Resistive elements are used here as analog memory with multiple levels of resistance in a single cell. Various types of resistive elements, such as oxide memristors [36,37], floating gate transistors (“flash”) [38,39], spintronic devices [40,41], have been proposed for neural networks.

In inference, the weights are not modified, therefore the characteristics of switching resistive memories are not relevant, but only their on and off resistances are. We assume characteristic on- and off-resistances in various resistive memory cells, Table 2. The parameters contributed from the memory cell per se are

$$I_{on} = V_{cc} / R_{on} \tag{51}$$

$$I_{off} = V_{cc} / R_{off} \tag{52}$$

$$a_{syn} = a_{dev} \tag{53}$$

The intrinsic capacitance of the synapse is of the order of that in a minimum interconnect, and the delay of synapses is determined by the upper bound of synapse resistance set at

$$R_{eff} = R_{on}\sqrt{n_l} \quad (54)$$

so

$$\tau_{syn} = 2.3 R_{eff} C_{ic} \quad (55)$$

$$E_{syn} = I_{on} V_{cc} \tau_{syn} \quad (56)$$

Another contribution comes from interconnects in the core and is described in Section 5.

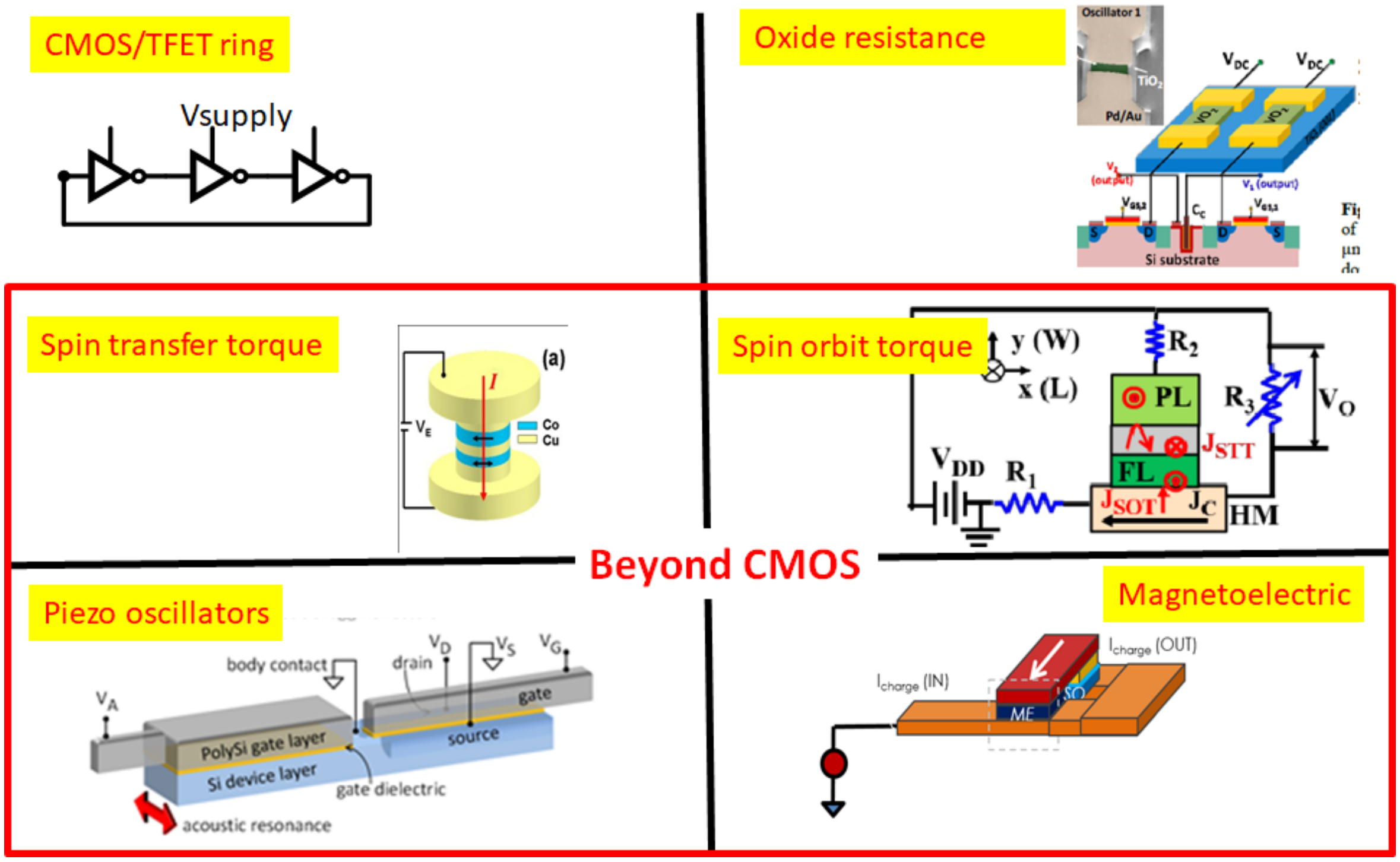

Figure 4. Schemes of oscillators used as synapses.

## 4. Types of Neural Networks

We classify neural networks into 4 types according to the nature of signals used, Figure 5.

a) Artificial neural network (ANN) where outputs switch in response to inputs in a mostly monotonic fashion.

b) Cellular neural network (CeNN) differ from ANN by their rectangular grid geometry and high connectivity. Here they are treated in agreement with [22] based on the methodology in [9].

c) Spiking neural network (SNN) receive trains of spikes at inputs. Synapses route spikes towards neurons, and neurons fire output spikes depending on their timing.

d) Oscillator neural network (ONN) determine the degree of pattern matching from the synchronization of oscillators in the array.

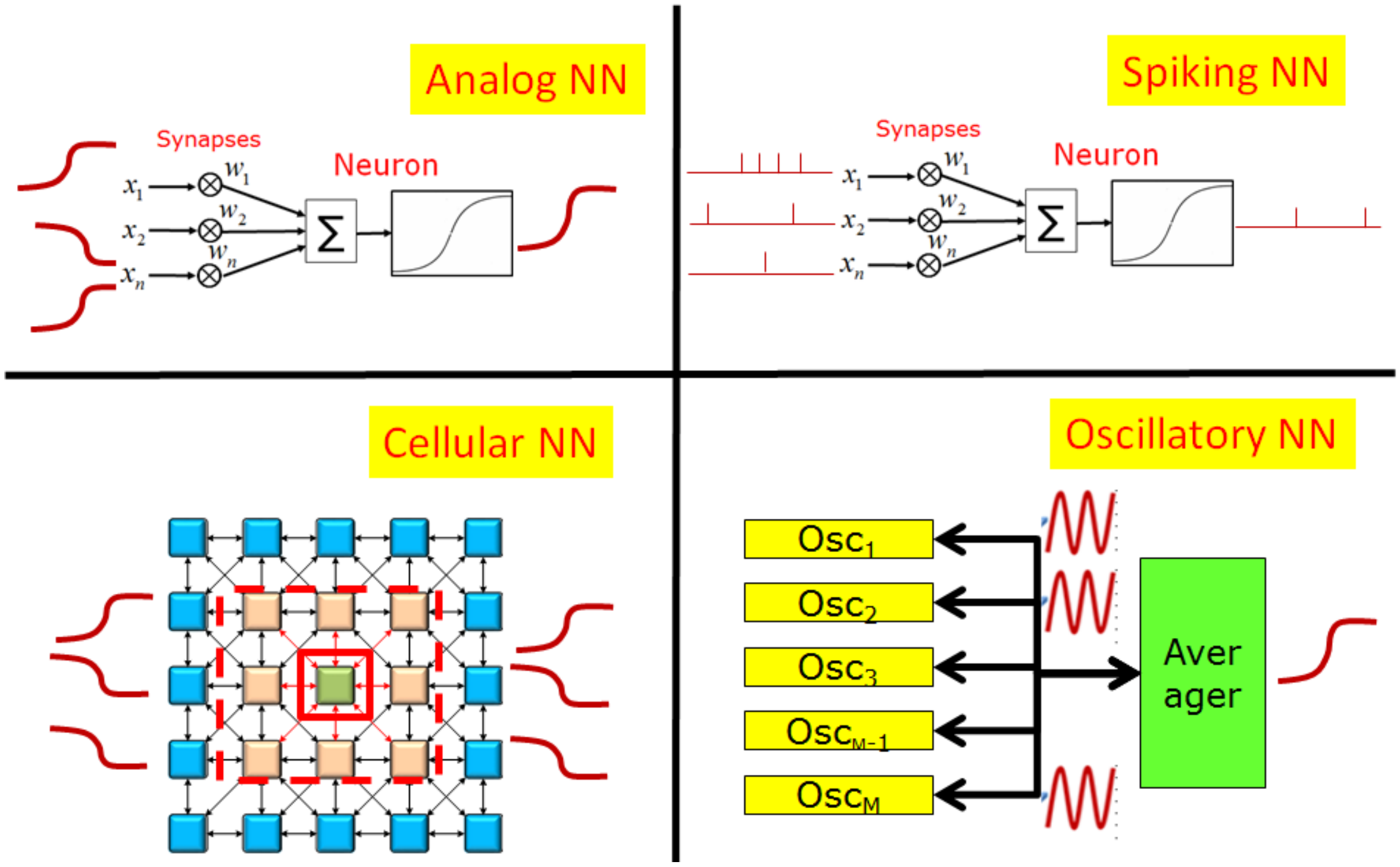

Figure 5. Schemes of the four types of neural networks considered in this paper.

Neural networks can be created with various combinations of neurons and synapses, Table 3. The first three letters of the label ("ANN","CNN","Spi",'Osc") designate the type of a neural network, the next set of letters (up to three) designates the type of neurons, and the last set of letters (up to four) designates the type of synapses. In neural networks, the synapse and neuron circuits require tens of transistors. Alternatively, single spintronic devices are capable of implementing synapses and neurons [27].

Table 3. LABELS FOR DEVICES/ARCHITECTURE COMBINATIONS

| Neuron | Synapse | ANN, CNN, SNN + ... | ONN |
|---|---|---|---|
| Digital CMOS | Digital CMOS 6T SRAM | DiCSRAM | |
| Digital CMOS | Digital CMOS MAC | DiCCMAC | |
| Digital CMOS | Oxide memristor digital | DiCOxme | |
| Digital TFET | Digital TFET MAC | DiTTMAC | |
| Digital CMOS | FEFET digital | DiCFETb | |
| Digital CMOS | Spin-transfer torque digital | DiCSTTb | OscSTT |
| Digital CMOS | Spin-orbit digital | DiCSOTb | OscSOT |
| Analog CMOS | Analog CMOS | AnCAnC | OscMOSring |
| Analog TFET | Analog TFET | AnTAnT | OscTFEring |
| Analog CMOS | Ferroelectric FET | AnCFET | OscPiezo |
| Analog CMOS | Oxide memristor | AnCOxme | OscOxide |
| Analog CMOS | Floating gate | AnCFlGa | |
| Analog CMOS | PCM | AnCPCM | |
| Ferroelectric FET | Ferroelectric FET | FETFET | |
| Domain wall | Domain wall | DoWDoW | |
| Spin-orbit torque | Spin-orbit analog | SOTSOTa | |
| Magnetoelectric | Magnetoelectric | MEME | OscME |

We will adopt the same synapses across the ANN, CeNN, and SNN classes, though neurons will be different.

<u>ANN.</u>
This is the default case, we directly use the estimates for the synapses and neurons obtained in the previous section.

<u>CeNN.</u>
We follow the treatment of cellular neural networks in [22]. Application of CeNN to CoNN was considered in [42]. Due to both feedback and feedforward connections in a CeNN and due to more connections than just nearest neighbors, the number of synapses is doubled. Also it takes a longer time for CoNN networks to settle to the steady state due to a larger number of connections [22]. This delay depends on the input patterns; we take estimated average values. Therefore

$$a_{syn} = M_{syncnn} a_{syn,ann} \tag{57}$$

$$\tau_{syn} = M_{stepcnn} M_{syncnn} \tau_{syn,ann} \tag{58}$$

$$E_{syn} = M_{stepcnn} M_{syncnn} E_{syn,ann} \tag{59}$$

$$a_{neu} = a_{neu,ann} \tag{60}$$

$$\tau_{neu} = M_{stepcnn} \tau_{neu,ann} \tag{61}$$

$$E_{neu} = M_{stepcnn} E_{neu,ann} \tag{62}$$

Neural network parameters related to Hebbian learning are based on the synaptic weight information: the maximum weight value obtained from the training weights, and the average summation of the weights per cellular cell [22], Table 1.

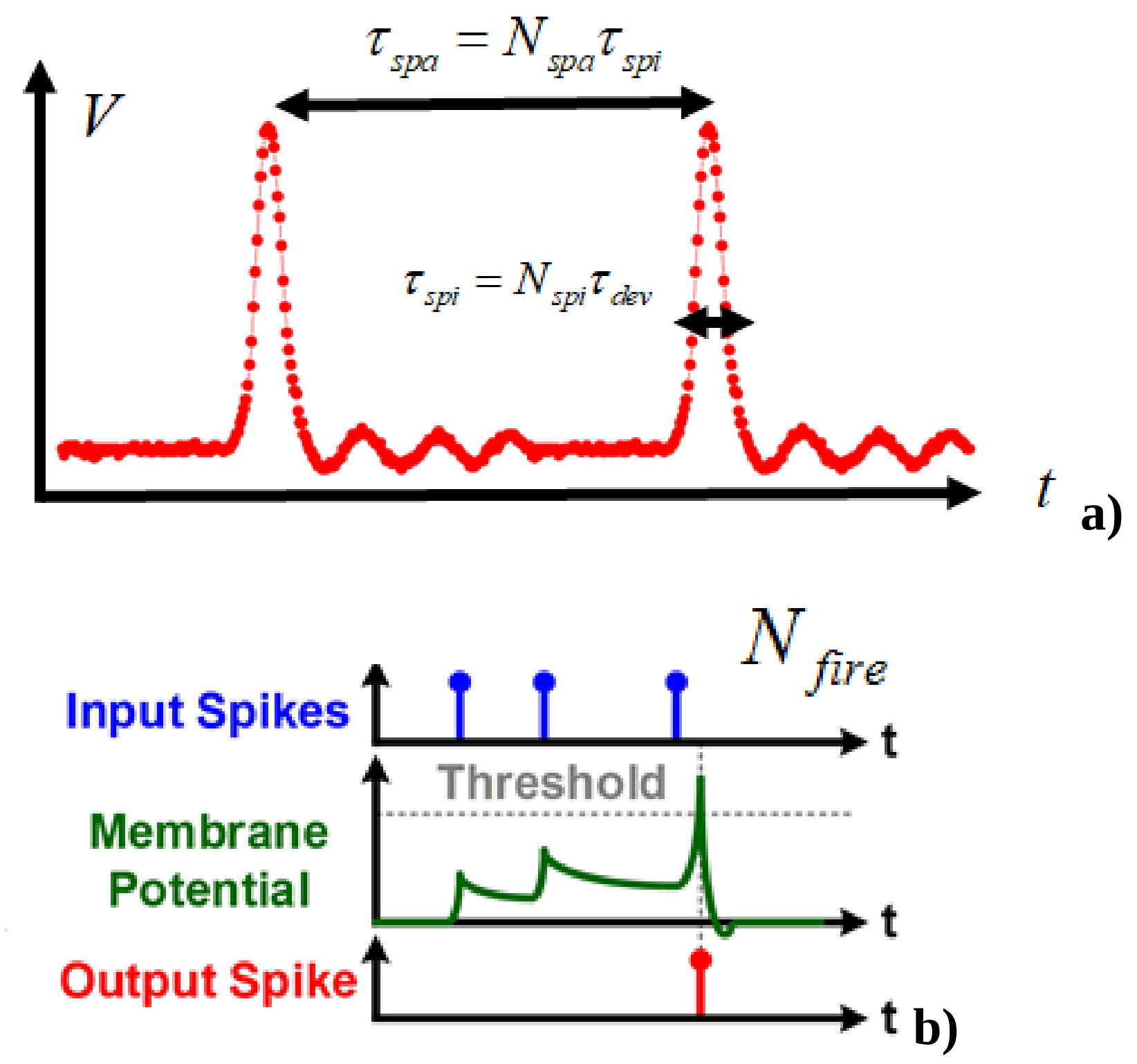

Figure 6. Approximate wave forms in a spiking neural network. a) The spike separation is longer than the spike duration; b) Multiple synapse spikes are required for a neuron to fire [65].

SNN.

We introduce a factors (Table 1) relating the spike duration to the device delay and relating the time spacing between spikes to the spike duration, Figure 6.

With these factors the estimates for SNN become

$$\tau_{syn} = \tau_{syn,ann} N_{spi} N_{spa} \quad (63)$$

$$E_{syn} = E_{syn,ann} N_{spi} \quad (64)$$

$$\tau_{neu} = \tau_{neu,ann} N_{spi} N_{spa} N_{fire} \quad (65)$$

$$E_{neu} = E_{neu,ann} N_{spi} N_{fire} \quad \text{(rate coded)} \quad (66)$$

$$E_{neu} = E_{neu,ann} N_{spi} \quad \text{(temporal coded)} \quad (67)$$

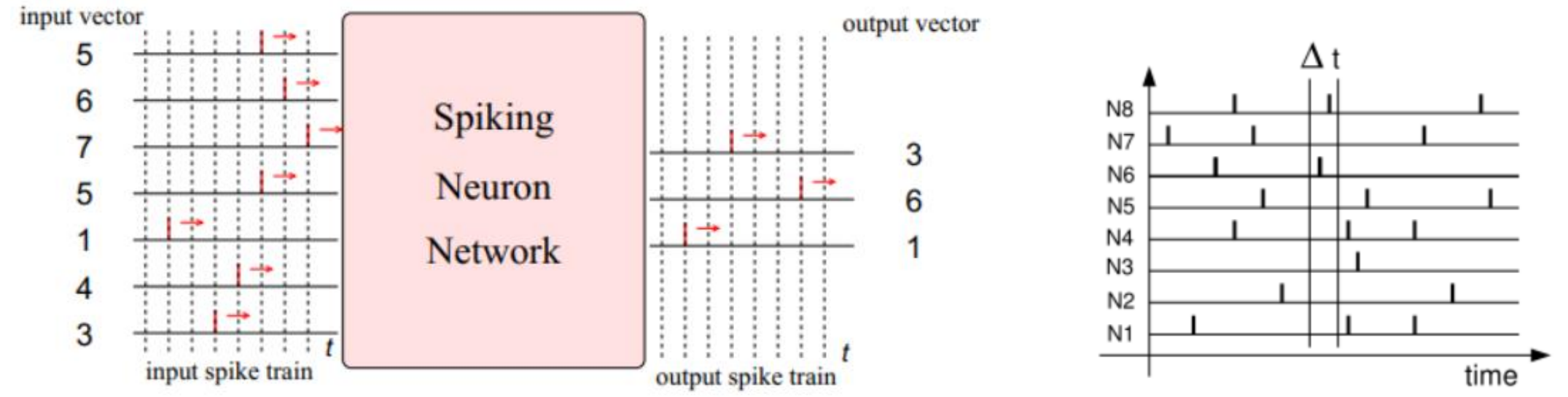

Figure 7. Two types of spiking NN: rate coded and temporal coded.

Note that it takes a different number of spikes arriving at a neuron from synapses to make it fire for the cases of rate coding or temporal coding of the signal, Figure 7. We also account for the spiking activity, i.e., the probability of a synapse producing a spike in a given spiking interval. We incorporate an empirical trend that the spiking activity decreases in the later stages where spike activity in an SNN decreases by $r_a = 1/n_{stage}$ with stage number in a DNN or CoNN [43].

<u>ONN.</u>

The area of oscillators is typically larger because they contain multiple instances of simple gates.

$$a_{syn} = 10a_{syn,ann} \quad (68)$$

$$a_{neu} = 30a_{neu,ann} \quad (69)$$

The frequency of transistor-based ring oscillators is determined by the product of the number of inverters (chosen here to be 5) and an average delay in an inverter. The frequency of spintronic oscillators empirically proves to be several times faster than the inverse switching time of a logic device. The average power is proportional to that of a logic device.

$$f_{osc} = 0.1/\tau_{inv4} \quad \text{(for transistor oscillators)} \quad (70)$$

$$P_{osc} = 3E_{\text{int}} / \tau_{\text{int}} \quad \text{(for transistor oscillators)} \quad (71)$$

$$f_{osc} = 6/\tau_{neu,ann} \quad \text{(for spintronic oscillators)} \quad (72)$$

$$P_{osc} = 6E_{neu,ann} / \tau_{neu,ann} \quad \text{(for spintronic oscillators)} \quad (73)$$

$$f_{osc} = 1/\tau_{neu,ann} \quad \text{(for piezo oscillators)} \quad (74)$$

$$P_{osc} = 3E_{neu,ann} / \tau_{neu,ann} \quad \text{(for piezo oscillators)} \quad (75)$$

The operation of the ONN synapse is limited by the synchronization time of the oscillators which takes several periods of oscillations, Table 1. Thus the ONN benchmarks are

$$\tau_{syn} = N_{synch} / f_{osc} \tag{76}$$

$$E_{syn} = P_{osc}\tau_{syn} \tag{77}$$

$$\tau_{neu} = \tau_{syn} \tag{78}$$

$$E_{neu} = E_{syn} \tag{79}$$

## 5. Treatment of interconnects

The benchmarks for neural network elements, neural gates, and larger DNNs are built up hierarchically, from benchmarks for a synapse and a neuron obtained in the previous section. We refer to it as 'bottoms-up benchmarking'.

The chip comprises a number of neural cores with multiple neurons in each and multiple synapses feeding signals into each core. The total number of synapses per chip is thus

$$s_{ch} = c_{ch} n_{cor} s_{neu} \tag{80}$$

Empirical factors are introduced to account for layout overhead: spacing between circuits for interconnects, routing circuits, intermediate registers, encoders/decoders etc. To obtain the corrected area, the estimated area is multiplied by additional layout overhead factors, Table I. For a certain workload only a share of synapses $r_a$ may be active. The area of the chip is then

$$a_{ch} = M_{ch} c_{ch} \left( M_{cor} n_{cor} \left( M_{neu} a_{neu} + s_{neu} M_{syn} a_{syn} \right) \right) \tag{81}$$

The operation of an interconnect is mainly determined by the interconnect capacitance per unit length. The energy to charge an interconnect is

$$E_{ic} = c_{ic} l V^2 \tag{82}$$

where the length of an interconnect from a circuit block to the next block is calculated as

$$l = \sqrt{a_{circ}} \tag{83}$$

The area of the relevant circuit block: for synapses $a_{circ,syn} = a_{syn} s_{cor}$, set by the requirement to deliver the synapse output within the area of the core; for neurons $a_{circ,neu} = a_{ch}$, set by the requirement to deliver the output signal of a neuron to any part of the chip.

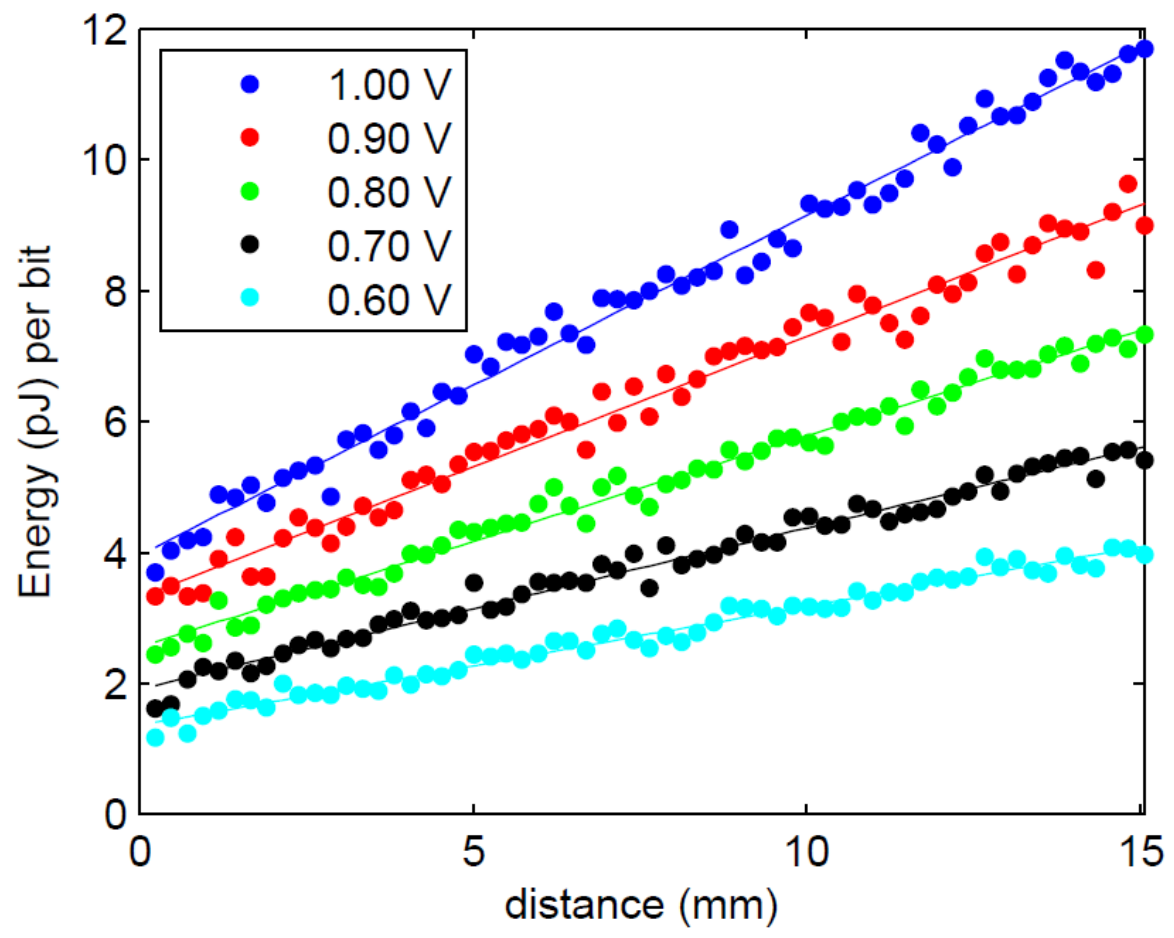

Figure 8. Energy per bit vs. distance in TrueNorth [4].

A geometry calculation with a low-k interlayer dielectric results in $c_{ic} = 10^{-10} F/m$ for 20nm wire width. Energy and capacitance vs. distance for an actual NN chip, TrueNorth [4], is shown in Figure 8. With voltage of 1V, the energy of a spike is 8pJ for 15mm of interconnect length, which implies that $c_{ic} = 5 \cdot 10^{-10} F/m$. Therefore the energy to transmit a bit over an interconnect in neural networks is less efficient by the factor of 5 than the energy of the ideal case, i.e. just charging the interconnect capacitance. This empirical factor of 5 is incorporated into further estimates.

The delay in a core-wide interconnect is dominated by the RC-delay in wires connecting synapses and neurons:

$$\tau_{cic} = \left(0.38 R_{ic} C_{ic} + R_{eff} C_{ic} + R_{ic} C_{load}\right) l / l_{ic} . \qquad (84)$$

The delay of charging a global, chip-wide interconnect

$$\tau_{gic} = \frac{c_{ic} l V}{I_{neu}} = \frac{E_{ic}}{I_{neu} V} . \qquad (85)$$

The delay and energy of a core-wide interconnect are added to those of a synapse. The energy and delay of a chip-wide interconnect are added to those of a neuron.

## 6. Chip-level benchmarks

The operation of the chip involves signals coming from input neurons, processed in synapses, and then firing of output neurons. A synaptic operation (synaptic event) is understood in non-spiking networks as an operation of multiplication of an input signal by a weight, i.e., “multiply-and-accumulate” (MAC). However in spiking networks, a synaptic even is mostly understood as “A spike event is a synaptic operation evoked when one action potential is transmitted through one synapse" according to [44]. There may be multiple spikes required to fire a neuron, i.e. multiple synaptic operation correspond to one MAC. Therefore the firing rate is conventionally defined differently for spiking, SNN, (to keep it consistent with definitions in [4]) and non-spiking, ANN, CeNN, and ONN, networks:

$$f_{fire} = 1/\tau_{syn} \quad \text{(non-spiking)} \tag{86}$$

$$f_{fire} = 1/\left(r_a s_{neu} \tau_{syn}\right) \quad \text{(spiking)} \tag{87}$$

The time step in a chip corresponds to the operation of one stage of a neural network. It consists of the time for enough synaptic inputs to arrive at the neuron to make it fire plus the delay in the neuron itself.

$$\tau_{step} = 1/f_{fire} + \tau_{neu} \tag{88}$$

Total energy per synaptic even contributed by a synapse and a share of a neuron energy

$$E_{syntot} = E_{syn} + E_{neu}/\left(r_a s_{neu}\right) \tag{89}$$

Publications normally quote the **throughput of synaptic operations** per second (SOPS) (not to be confused with throughput of inferences, see below)

$$T_{syn} = f_{fire} r_a s_{ch} \tag{90}$$

The dissipated power is

$$P_{ch} = T_{syn} E_{syntot} \tag{91}$$

Then the energy per time step is

$$E_{ch} = P_{ch} \tau_{ch} \tag{92}$$

To compare benchmarks between actual chips and bottoms-up estimates for various neural networks, we calculate the performance of the latter for a nominal chip with parameters, Table 1.

## 7. Neuromorphic computing workloads vs. hardware

We considered examples of neuromorphic workloads, including

1) CoNN such as LeNet [45,46], shown in Figure 9;
2) AlexNet [47];
3) a single stage convolution of a 35x35 pixel image with 24 filters of 5x5 pixels;
4) a single stage associative memory of pixel patterns [22];
5) a DNN for recognition of hand-written digits from the MNIST hand-written-digit image database [48] implemented as a multi-layer perceptron (MLP) with 784×256×128×10 fully-connected neurons in layers;
6) a DNN for speech recognition from [18] – a 4 layer MLP with 390x256x256x29 neurons.

While all of these networks belong to the class of non-recurrent DNN, these are examples of ubiquitous applications required by users. However these workloads may not be favorable to SNNs. For example they do not utilize temporal information carried by spikes. The reader should be warned that conclusions may change if we consider workloads more favorable to SNNs.

Now we determine the benchmarks for a part of the chip necessary to perform a specific computing workload (we will use the subscript CW). More specifically in our case the computing workload is an inference. It's hardware implementation is determined by the logic structure of a neural network, which can be thought of as an algorithm. Each feature map in a stage is produced by a convolution with one of the kernels; this process is mapped to a neural core. The number of neurons and synapses in each core is determined by the connectivity of the neural network. Each core will have a number of input neurons $n_{in}$ and a number of output neurons $n_{out}$. Often overlooked input neurons are shown in the array schemes, e.g. in [4]. Each output neuron collects inputs from the number of active synapses per neuron $s_{neu}$. For example, the LeNet NN shown in Figure 9 can be implemented by an application specific design comprising a set of neural cores shown in Figure 10. A general purpose neural chip, is composed of cores of a fixed size with some of the input and output neurons remaining unused.

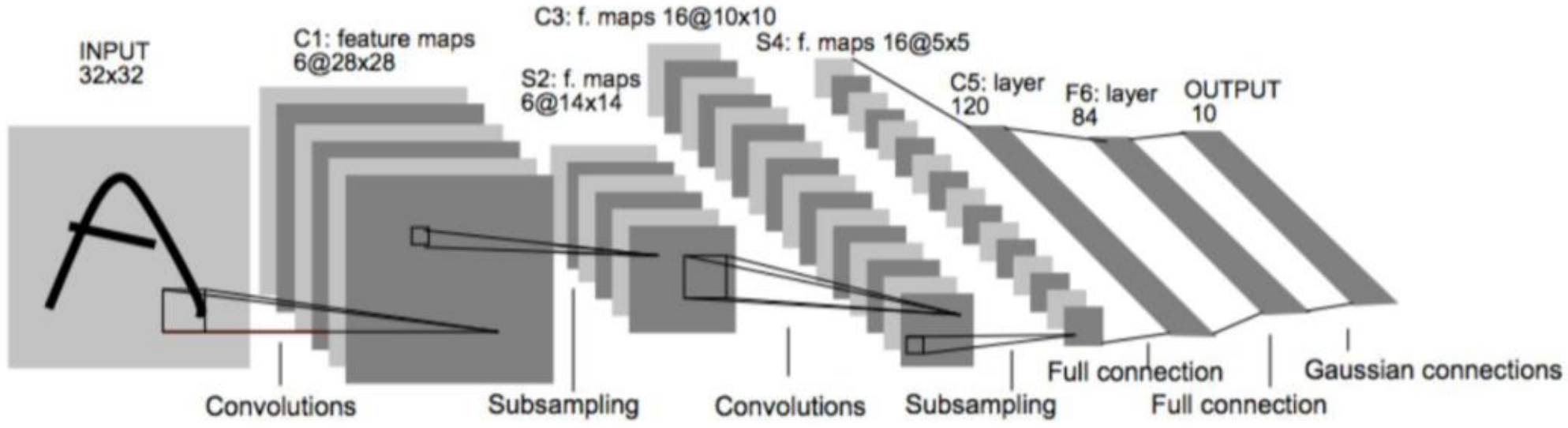

Figure 9. Map of layers in the CoNN, LeNet [45].

The total number of synapses in the deep neural network can be very large. It is much larger than the number of trainable weights (which can be e.g. filters for the feature maps in CoNN). In this benchmark we adopt an approach of multiple copies of weights written into the memory of synapses and placing them close to neurons. This requires a larger number of memory cells and the corresponding chip area dedicated to them. However it can be affordable in the case of dense analog memory. During training this requires more time and energy for updating the weights. The alternative – using only the necessary memory to hold trainable parameters – has the major downside ofthe need to route connections with numerous neurons and the time and energy to fetch the weight values.

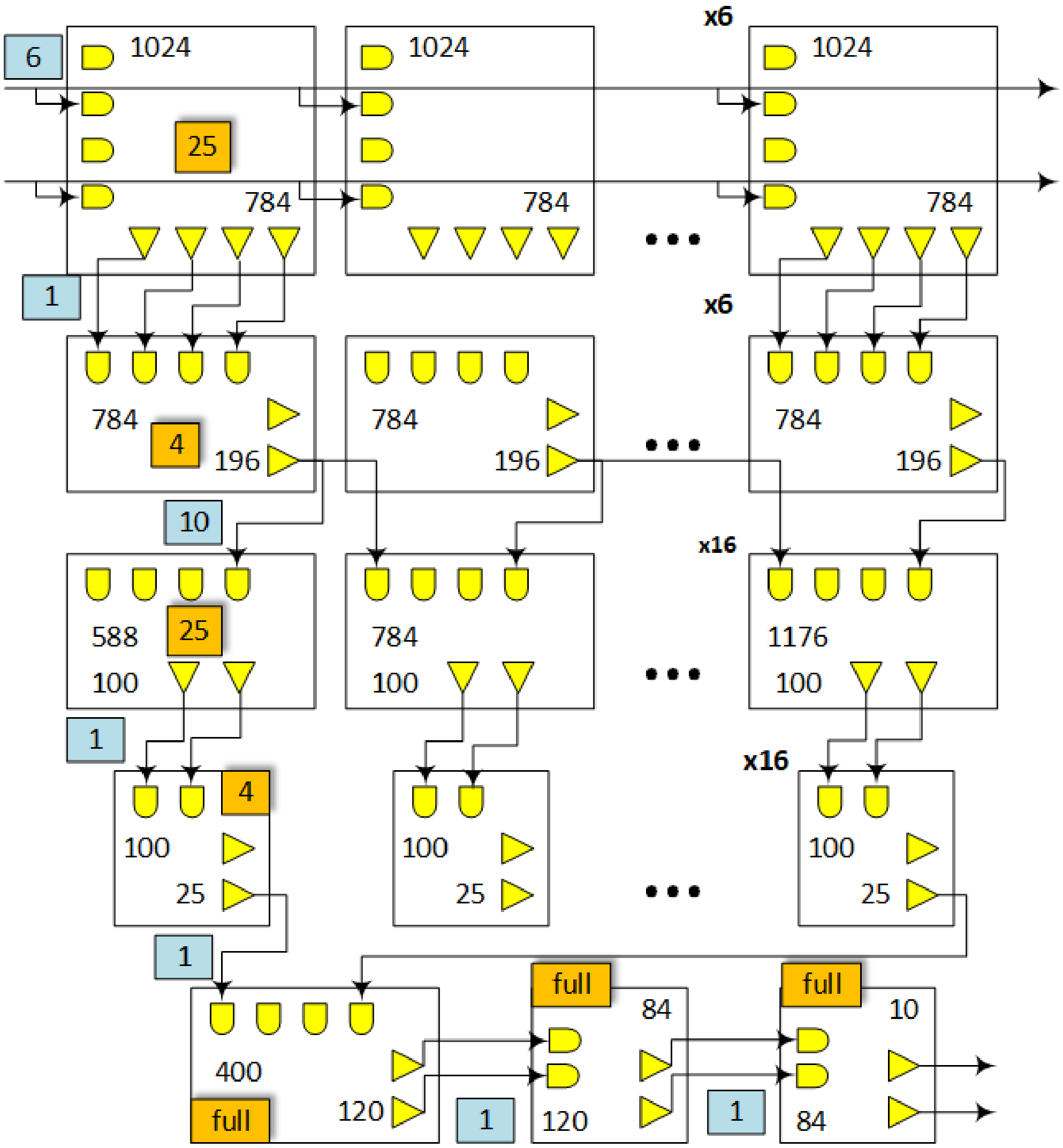

Figure 10. Block-diagram of an implementation of the CoNN, LeNet. "AND" symbols designate input neurons, triangles designate output neurons; their number of instantiations are indicated. Numbers in orange squares designate the number of synapses per neuron in the respective core; 'full' means a fully connected core. Numbers in blue squares next to buses connecting cores designate the fanout $f_o$ for output neurons.

Focusing on the structure of a core, we envision that different topologies can be chosen for interconnecting input and output neurons by synapses. The most straightforward one is the cross-connect (Figure 11). It is best for fully connected layers (such as those in the bottom part of Figure 10), and allows for a general pattern of connections. But it will leave many unused synapses in case of a sparsely connected NN. Since the processing of information is happening in or on the periphery of the memory array, containing the weights, such a scheme can also be classified as "compute-in-memory" or inference-in-memory".

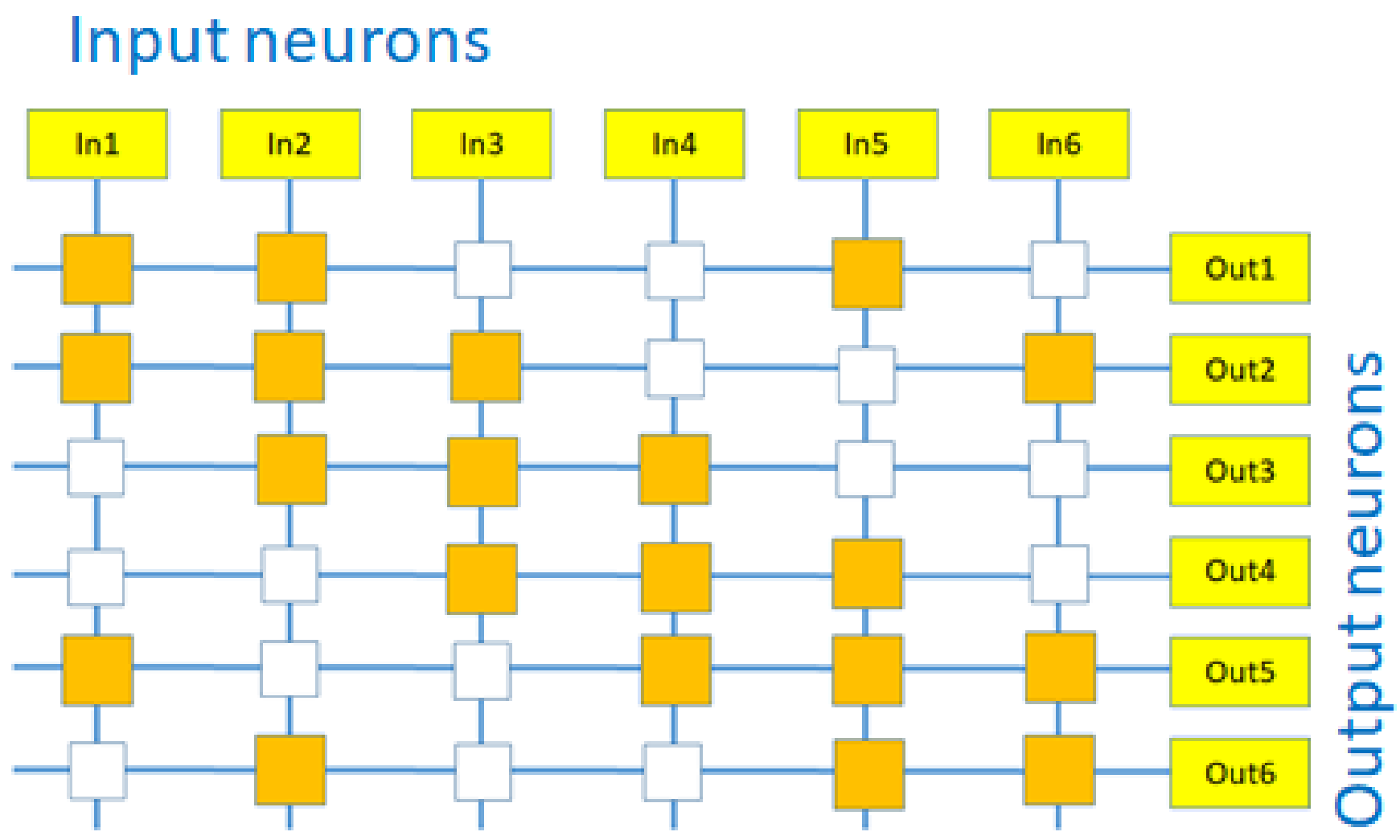

Figure 11. Cross-connect topology for the neural network. Input ('In') and output ('Out') neurons are shown in yellow. Active synapses are shown in orange, and unused synapses in white.

The convolution topology (Figure 12) is specifically designed for connections in convolution layers (such as those in the top part of Figure 10). This scheme utilizes the property of CoNN – sparse connectivity between neurons. It also closely resembles Cellular NN (CeNN) connectivity. In this topology the output neurons are placed close to connected input ones. They mimic the positions of pixels in the image and the resulting feature map. The convolution topology is efficient since it contains only active and no unused synapses. But it is not general – additional synapses need to be designed for a less sparse connectivity.

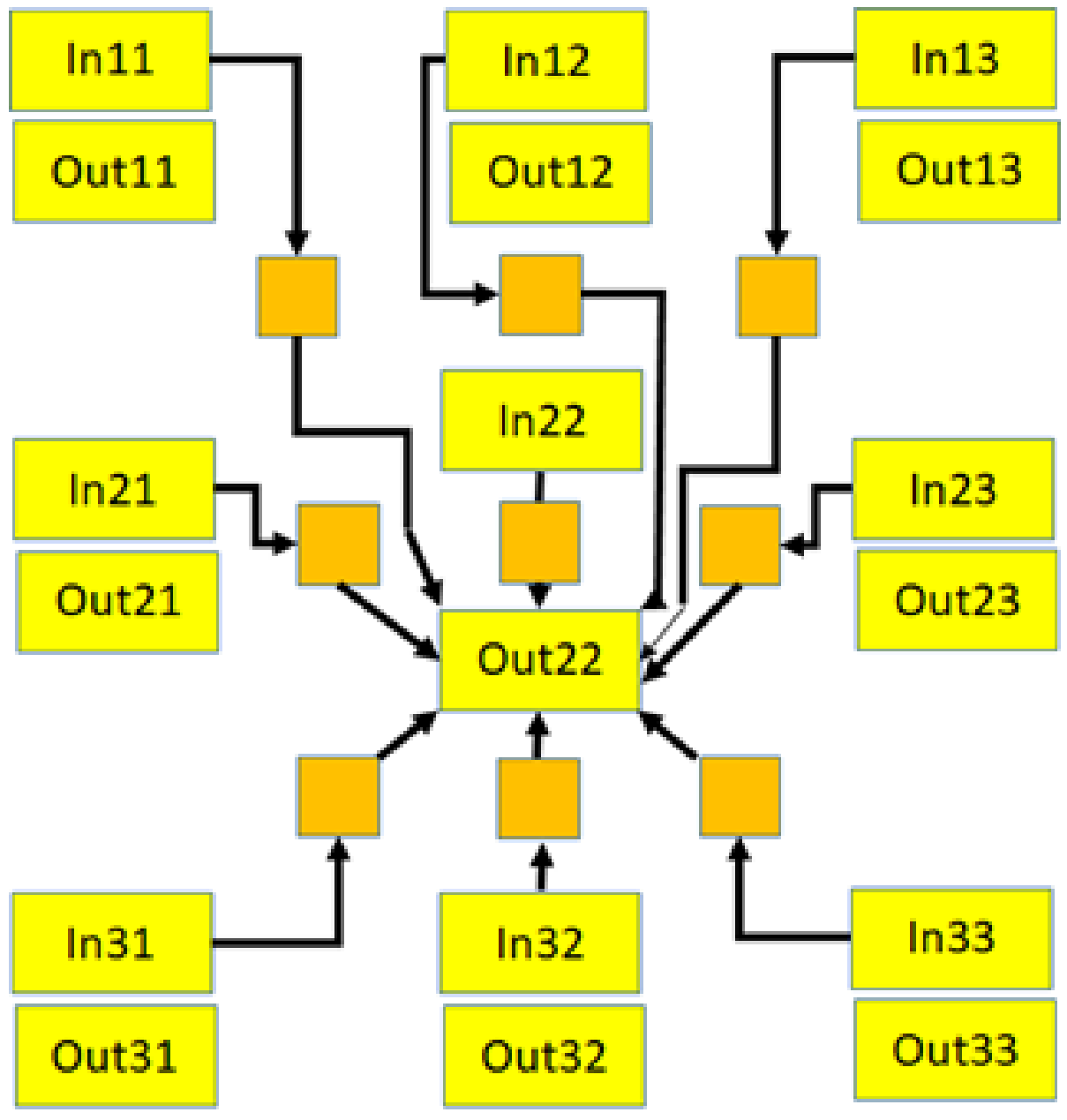

Figure 12. Convolution topology for the neural network. Input ('In') and output ('Out') neurons are shown in yellow. Active synapses are shown in orange.

The means by which long interconnects can be routed is shown in Figure 13. For the cross-connect topology, the routing is trivial, since both input and output neurons are at the edge of the core. Even for the convolution topology, the input and output wires can enter in a regular array of wires and still be routed to neurons. The pitch of interconnect wires is assumed to be $p = 8F$. Then the interconnect-wire-limited area of a core is

$$a_{wire} = n_{in} n_{out} p^2 \tag{93}$$

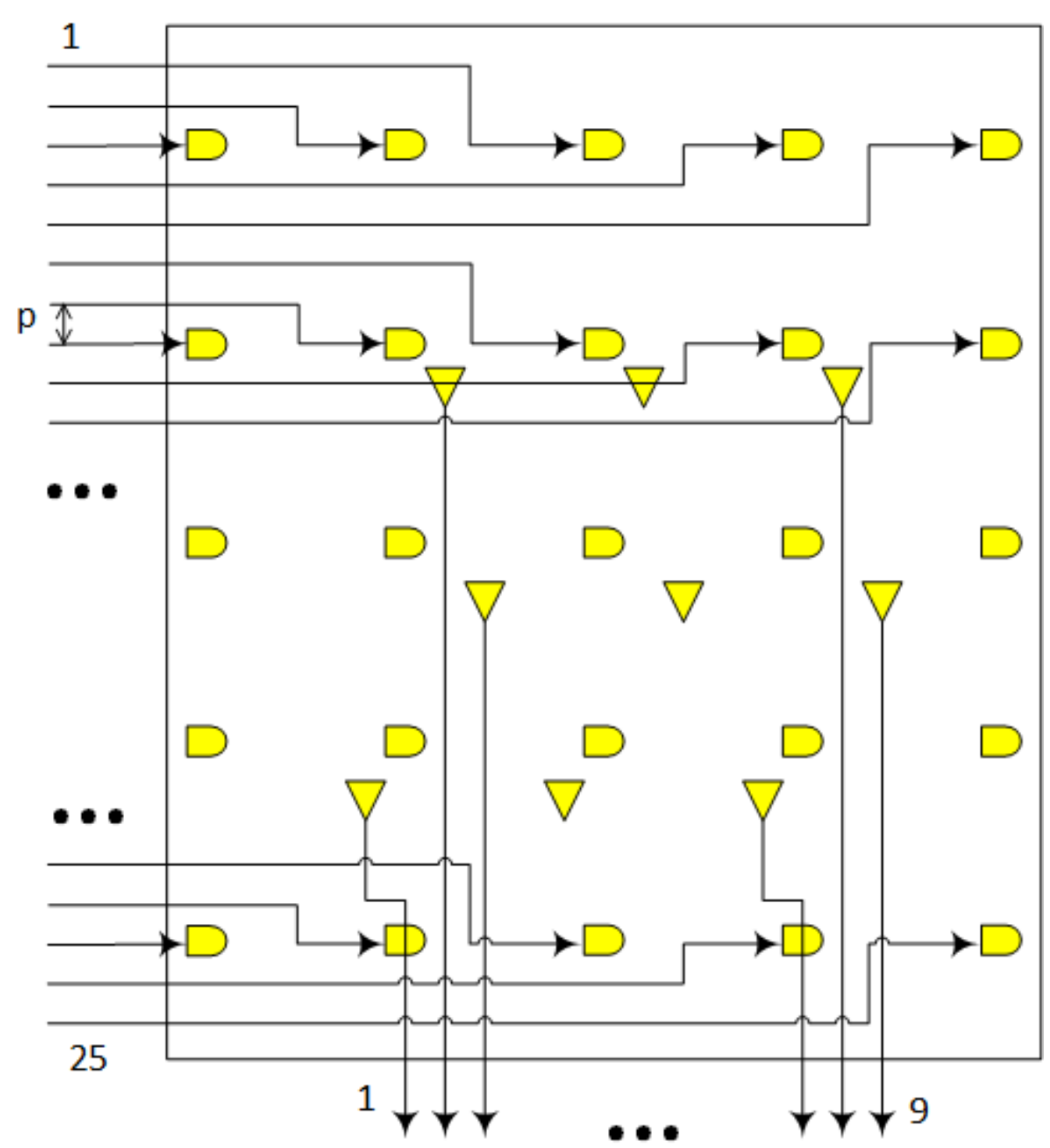

Figure 13. Inter-core interconnects and neurons for the convolution topology of a neural network. Pitch *p* is marked.

The speed of a neural gate is determined by its fan-in, i.e. the number of synaptic operations connected to one output neuron that can be performed in parallel. We will assume the fan-in of digital CMOS to be 2, analog CMOS to be 16, spintronic devices to be 32. The fan-in for spiking NN is assumed unlimited regardless of a device. We will apply this limitation to bottoms up benchmarks, and consider it satisfied for tops-down benchmarks.

In the case when the devices forming neurons have a limited fan-in $f_i$, they need to be cascaded, as shown in Figure 14. The number of levels of cascading is (rounded to next higher integer)

$$l_{cas} = ceil\left(\log_{fi} s_{neu}\right) \quad (94)$$

Then the number of neurons to form a fan-in in a neural gate is

$$n_{cas} = \left(f_i^{lcas} - 1\right) / \left(f_i - 1\right) \quad (95)$$

so that the number of neurons per core is

$$n_{cor} = n_{cas} n_{out} + n_{in} \quad (96)$$

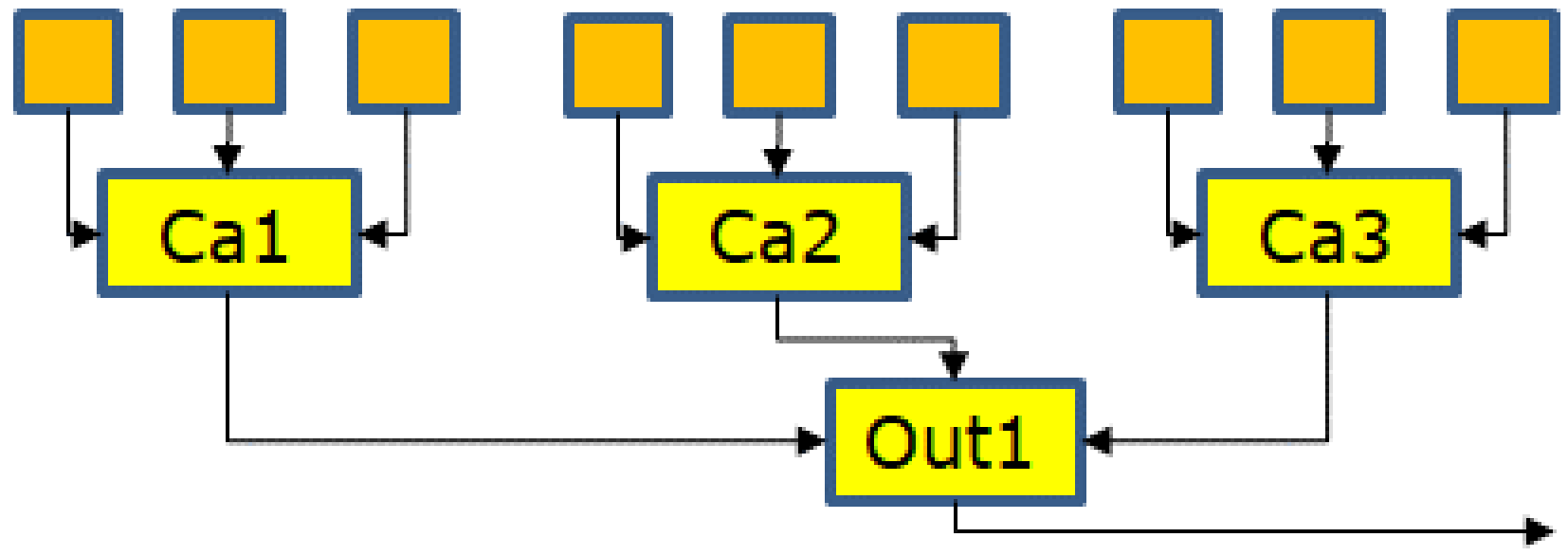

Figure 14. Synapses connecting to the output neuron via cascaded neurons in case of a limited fan-in.

In the cross-connect case, the area of a core (performing a stage of a NN) is (provided that $n_{in}$ is larger than $s_{neu}$ ).

$$a_{cor} = M_{cor}\left(M_{neu} a_{neu} n_{cor} + M_{syn} a_{syn} n_{out} n_{in}\right) \quad (97)$$

In the convolution case

$$a_{cor} = M_{cor}\left(M_{neu} a_{neu} n_{cor} + M_{syn} a_{syn} n_{out} s_{neu}\right) \quad (98)$$

We then take the larger of this estimate and the interconnect wire limit $a_{wire}$ to constitute the area of a core $a_{st}$ for the given stage of CoNN. The time and energy for a stage of the computing workload is

$$\tau_{st} = l_{cas} \tau_{syn} + \tau_{neu} \quad (99)$$

$$E_{st} = r_a s_{neu} n_{out} E_{syn} + n_{out} E_{neu} \quad (100)$$

If the fan-in of the devices is small, which makes its cascading impractical, the synaptic operations can be performed sequentially. We will be using this method for neural accelerators. In this case the fan-in factors above are set to 1, and instead the time estimate changes to

$$\tau_{st} = s_{neu}\tau_{syn} + \tau_{neu} \tag{101}$$

In a multi-stage network, like in Figure 9, each stage is using one core. Therefore the above benchmarks need to be multiplied by the number of feature maps in each stage and then summed over all stages to obtain benchmarks for a computing workload.

$$a_{CW} = \sum_{st} a_{st} f_{st} \tag{102}$$

$$\tau_{CW} = \sum_{st} \tau_{st} \tag{103}$$

$$E_{CW} = \sum_{st} E_{st} f_{st} \tag{104}$$

In case where the implementation of CoNN is constrained by area, all the cores in Figure 9 can be replaced by a single core, and all the stage operations can be performed in a sequential ('time-multiplexed') manner. Here we neglect the energy and delay of storing the intermediate results. Then in this case the estimates need to change to

$$a_{CW} = \max_{st}\left(a_{st}\right) \tag{105}$$

$$\tau_{CW} = \sum_{st} \tau_{st} f_{st} \tag{106}$$

This is the case ,for example for all networks using a digital multiplier in a synapse (labeled "MAC"). This treatment is used for all tops-down estimates of implemented chips.

Then the power in a computing workload and the **throughput of inferences** (not to be confused with the synaptic throughput) in units of inferences per second (IPS) per unit area is

$$P_{CW} = E_{CW} / \tau_{CW} \tag{107}$$

$$T_I = 1/\left(a_{CW}\tau_{CW}\right) \tag{108}$$

## 8. Prototype neuromorphic chips

We will compare the above benchmarks with those for prototype chips fabricated and measured by several groups of researchers. In this chapter we consider mostly spiking neuromorphic chips [5,6,49]. To them we apply 'tops-down benchmarking', i.e. calculate the neuron and synapse values from the total number of synapses, the total area, and known synaptic throughput. A number of such chips have been previously benchmarked [50]. One should keep in mind the difference between the 'bottoms-up' and 'tops-down' benchmarks. The former are for idealized circuits and do not include multiple auxiliary circuits necessary for the operation of a chip. The latter are complete real-life chips and contain all the circuit overheads which are hard to quantify. However we will sometimes put these two types of benchmarks side-by-side for sanity checks and extract some insights, given the above caveats. We assume that 5% of the chip area is occupied by neurons and the rest by synapses. The area per neuron and per synapse is thus

$$a_{neu} \approx 0.05 a_{ch} / \left( c_{ch} n_{cor} \right) \qquad (109)$$

$$a_{syn} \approx 0.95 a_{ch} / \left( c_{ch} n_{cor} s_{neu} \right) \qquad (110)$$

Synaptic time step $\tau_{syn}$ is calculated from the firing rate $f_{fire}$ as per the previous section. Publications mostly quote the energy per synaptic even. We approximate the energy per neuron as

$$E_{neu} = E_{syn} r_a s_{neu} \qquad (111)$$

The synaptic throughput, power, and energy per time step is calculated like in Section 5. The input parameters from cited publications are collected in Table 4. In some cases the input parameters are not available, so they are calculated from the consistency of quoted synaptic throughput with that calculated from equations in Section 6. Then we use these inputs to obtain the benchmarks for a synapse and a neuron, and calculate benchmarks for various computing workloads described in Section 7.

Table 4. PARAMETERS FOR NEUROMORPHIC CHIPS

| Chip Name | Main Affiliation | Year | # cores | Neurons per core | Synapses per neuron | Area, mm² | Power, mW | Syn Throughput, MSOPS | Energy syn event, pJ | Syn fire rate, $s^{-1}$ | Activity | Process, nm | Voltage, V | References |
|---|---|---|---|---|---|---|---|---|---|---|---|---|---|---|
| Notation | | | $c_{ch}$ | $n_{cor}$ | $s_{neu}$ | $a_{ch}$ | $P_{ch}$ | $T_{syn}$ | $E_{spi}$ | $f_{syn}$ | $r_a$ | | | |
| **HICANN** | Heidelberg | 2010 | 1 | 512 | 224 | 50 | 1150* | 11,500 | 100 | 100k | 1 | 180 | 1.8 | [51] |
| **HICANN-X** | Heidelberg | 2018 | 1 | 512 | 256 | 32 | 2100* | 2600 | 800 | 20k | 1 | 65 | 1.2 | [52] |
| **SyNAPSE** | HRL | 2013 | 1 | 576 | 128 | 42 | 130 | 15 | 8700 | 203* | 1 | 90 | 1.4 | [53] |
| **SpiNNaker** | Manchester | 2013 | 16 | 1024 | 1024 | 102 | 1000 | 64 | 16k* | 10 | 0.4* | 130 | 1.2 | [54][55] |
| **SpiNNaker 2** | Manchester | 2017 | 64 | 2048 | 1024 | ? | 110 | 250 | 440 | 10 | 0.2* | 28 | 1.0 | [56] |
| **True North** | IBM | 2014 | 4096 | 256 | 256 | 430 | 72 | 3000 | 26 | 20 | 0.5 | 28 | 0.78 | [4][57] |
| **Neurogrid** | Stanford | 2014 | 1 | 65536 | 1024 | 168 | 59* | 62.5 | 941 | 10 | 0.09* | 180 | 1.8 | [58] |
| **IFAT** | UCSD | 2014 | 32 | 2048 | 1024 | 16 | 1.57 | 73 | 22 | 10 | 0.11* | 90 | 1.2 | [59] |
| **ROLLS** | ETH | 2015 | 1 | 256 | 512 | 51.4 | 4 | 4 | 1000* | 30 | 1 | 180 | 1.8 | [60][61] |
| **DYNAP-SEL** | ETH | 2016 | 4 | 256 | 64 | 43.8 | ? | ? | 50 | 30 | ? | 28 | 1.0 | [62] |
| **Loihi** | Intel | 2018 | 128 | 1024 | 128 | 60 | 450 | 30,000 | 15* | 1800* | 1 | 14 | 0.75 | [63][64] |
| **SBNN** | Intel | 2018 | 64 | 64 | 256 | 1.72 | 209 | 25,200 | 8.3 | 50k | 0.5* | 10 | 0.53 | [65] |

* derived value

We compared our performance estimates with experimentally measured [18] for the speech recognition workload on the Loihi and Mydiad2 (Movidius) chips. We note that the theoretical estimates are much more optimistic than experimental. The reasons for the discrepancy could be the circuit overhead required in an actual chip such as stand-by power, need to fetch the data, slower clock frequency, etc.

Table 5. COMPARISON OF BENCHMARKS WITH MEASURED PERFORMANCE

| | Loihi [18] | Loihi this work | Movidius [18] | Movidius this work |
|---|---|---|---|---|
| Speed, inference/s | 89.8 | 55k | 300 | 167k |
| Energy, μJ/inference | 770 | 6 | 1500 | 5.5 |

## 9. Digital Neural Accelerators

There is another type of chip being fabricated, which are commonly called neural accelerators [3]. They are based on traditional digital chips and in this sense are different from other neuromorphic hardware. Unlike CPU and GPU chips which implement neural network algorithms in software, neural accelerators have dedicated hardware engines to implement neural networks. They are highly optimized for vector-matrix multiplications, which is the core operation in neural algorithms. In this sense they present a good comparison for digital CMOS neural networks. The input data for them are collected in Table 6.

Our treatment of them is similar to that in Section 8, but adjusted for the non-spiking type. In this case the clock frequency determines the time step. Performance of these chips is often quoted in MAC/s. A MAC counts as two floating point operations (FLOP), multiplication and addition, although different delay and energy is required between the two of them. A significant share of the area of these chips is occupied by the cache, control circuits, etc. For this estimate we assume that 10% of the chip area is occupied by neurons and synapses. We use the inputs collected in Table 6, obtain the benchmarks for a synapse and a neuron, and then calculate benchmarks for various computing workloads described in Section 7.

Table 6. PARAMETERS FOR DIGITAL NEURAL ACCELERATORS

| Chip Name | Main Affiliation | Year | # cores | Neurons per core | Synapses per neuron | Memory Bytes | Area, mm² | Power, W | Performance, GMAC/s | Synapse energy, pJ | Clock frequency, MHz | Process, nm | References |
|---|---|---|---|---|---|---|---|---|---|---|---|---|---|
| Notation | | | $c_{ch}$ | $n_{cor}$ | $s_{neu}$ | $m_{ch}$ | $a_{ch}$ | $P_{ch}$ | $T_{syn}$ | $E_{syn}$ | $f_{cl}$ | | |
| **Diannao** | CAofS | 2014 | 1 | 16 | 16 | 2k | 3.02 | 0.485 | 452 | 1.1* | 980 | 65 | [66] |
| **Dadiannao** | CAofS | 2014 | 16 | 16 | 16 | 32M | 67.73 | 15.97 | 5585 | 2.9* | 606 | 28 | [67] |
| **Pudiannao** | CAofS | 2015 | 1 | 16 | 16 | 32k | 3.51 | 0.596 | 1056 | 0.56* | 1000 | 65 | [68] |
| **Shidiannao** | CAofS | 2015 | 1 | 16 | 16 | 36k | 4.86 | 0.32 | 194 | 1.7* | 1000 | 65 | [69] |
| **Eyeriss** | MIT | 2016 | 1 | 1 | 168 | 192k | 12.25 | 0.278 | 33.6 | 8.3* | 200 | 65 | [70] |
| **EIE** | Stanford | 2016 | 1 | 64 | 8 | 10.3M | 40.8 | 0.579 | 51.2 | 11.3* | 800 | 45 | [71] |
| **Origami** | ETH | 2016 | 1 | 4 | 49 | 43k | 3.09 | 0.654 | 98 | 6.7* | 500 | 65 | [72][73] |
| **Envision** | Leuven | 2017 | 1 | 16 | 16 | 128k | 1.87 | 0.044 | 51 | 0.86* | 200 | 28 | [74] |
| **TPU** | Google | 2017 | 1 | 256 | 256 | 28M | 300 | 40 | 11400 | 3.5* | 700 | 28 | [2] |
| **Tesla** | Nvidia | 2017 | 80 | 32 | 32 | 6M | 815 | 300 | 14900 | 20* | 1300 | 12 | [75] |
| **DPU** | Wave | 2018 | 16384 | 1 | 1 | 24M | 400 | 200 | 3900 | 51* | 6700 | 16 | [75] |
| **Q4MobilEye** | Intel | 2018 | 1 | 32 | 32 | 1M | ? | 3 | 1078 | 2.8* | 1000 | 28 | [75] |
| **Parker** | Nvidia | 2016 | 1 | 256 | 256 | 4M | ? | 5 | 375 | 13.3* | 3000 | 16 | [75] |
| **S32V234** | NXP | 2017 | 1 | 64 | 64 | 4M | ? | 5 | 512 | 9.8* | 1000 | 28 | [75] |
| **Myriad 2** | Intel | 2017 | 12 | 4 | 16 | 2M | 27 | 1.5 | 58 | 26* | 800 | 28 | [76] |

* derived value; ** ‘CAofS’ designates the Chinese Academy of Sciences.

## 10.Results for physical performance

The most informative view with the benchmarks is provide by the comparison of operation delay and energy. Such energy-delay plots are provided both for synapses (Figure 15) and for neurons (Figure 16). In many subsequent benchmarks the following technology options are found to be placed in close proximity to each other: the group of DiCFETb, DiCOxme, DiCSTTb, DiCSOTb and the group of AnCFET, AnCOxme, AnCFlGa, AnCPCM. In other words these are NNs with digital CMOS neurons (for the first group) or analog CMOS neurons (for the second group) which have very similar designs within each group. The difference within each group is the type of resistive memory comprising a set of binary bits (for the first group) or an analog resistive elements (for the second group). The analysis shows the kind of resistive memory produces noticeable but minor differences. In the following plots, for clarity, we will be suppressing the labels, leaving just one from each group.

One observes that among the four NN types, their neurons have similar ranges of energy. However on the average, the delay when ordered from fastest to slowest is as follows: ANN, ONN, CNN, and SNN. Within each type of NN, networks with both magnetoelectric neurons and synapses (MEME) show the lowest energy. This is in line with benchmarks for Boolean computing [9]. The networks with both ferroelectric neurons and synapses (FEFET) show the fastest speed. This is a result of the combination of relatively fast switching of a transistor and the assumption that only a single ferroelectric transistor is capable of performing the neuron function. The NNs based on a multiplier and adder in each synapse (MAC) prove to be the slowest and the most energy-consuming due to a large number of switching transistors in each such element. In general, NNs with analog neurons are faster and more energy efficient than similar NNs with digital neurons. This is due to the fact that the neural function is performed in parallel in analog NNs rather than synapse-sequential in digital NNs.

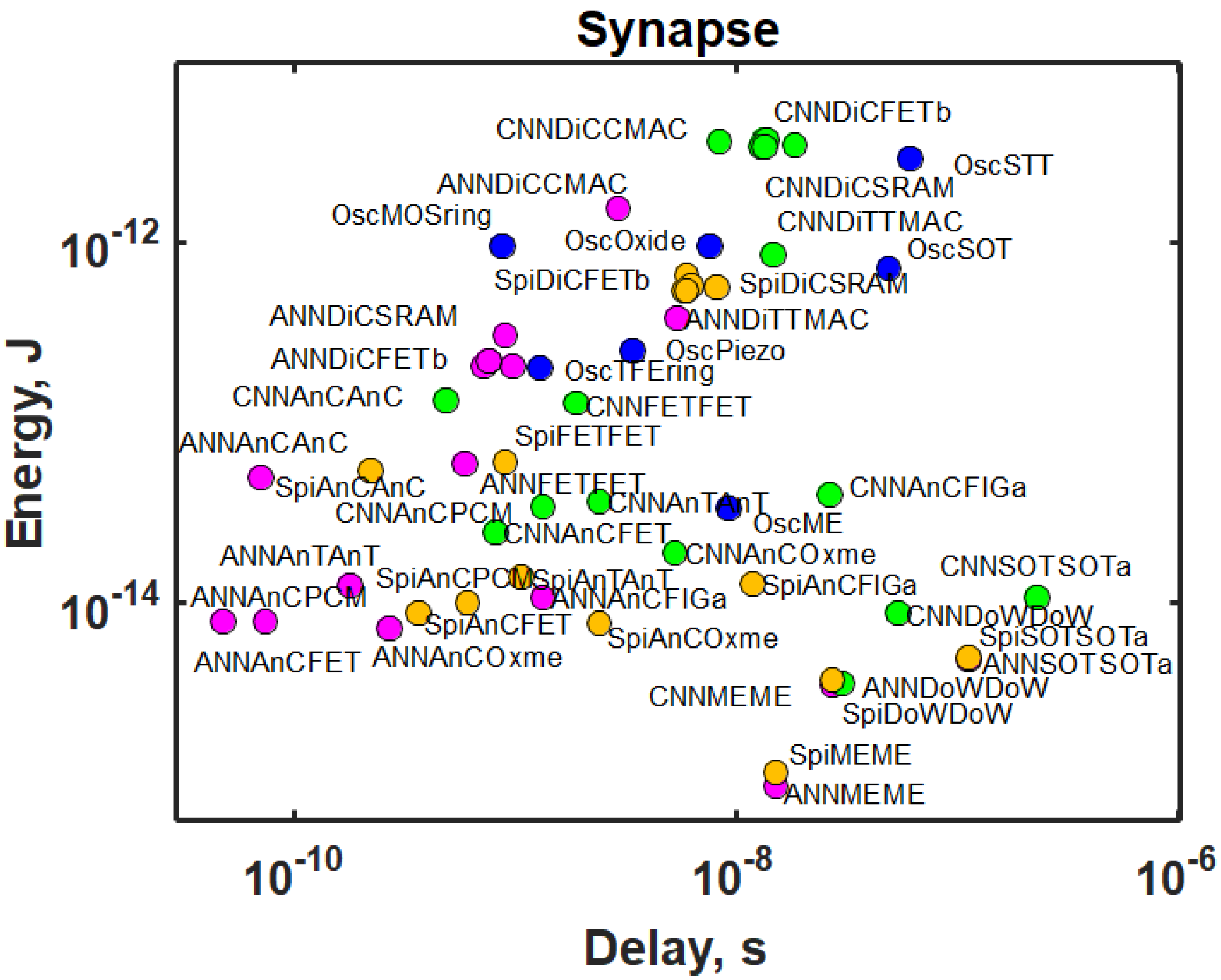

Figure 15. Energy vs. delay for a Synapse in ANN (magenta dots), CeNN (green dots), SNN (gold dots), ONN (blue dots). Labels for architectures according to Table 3.

The separation of four NN types is not so clear for synapses. The overall ranges of energy are similar between the types. However on the average, the delay ordered from fastest to slowest is as follows: ANN, SNN, CNN, and ONN. The difference with the similar relation in neurons is due to the fact that synapses are similar between ANN and SNN, and the difference arises in the operation on the core level. Other trends for synapses are similar to those for neurons.

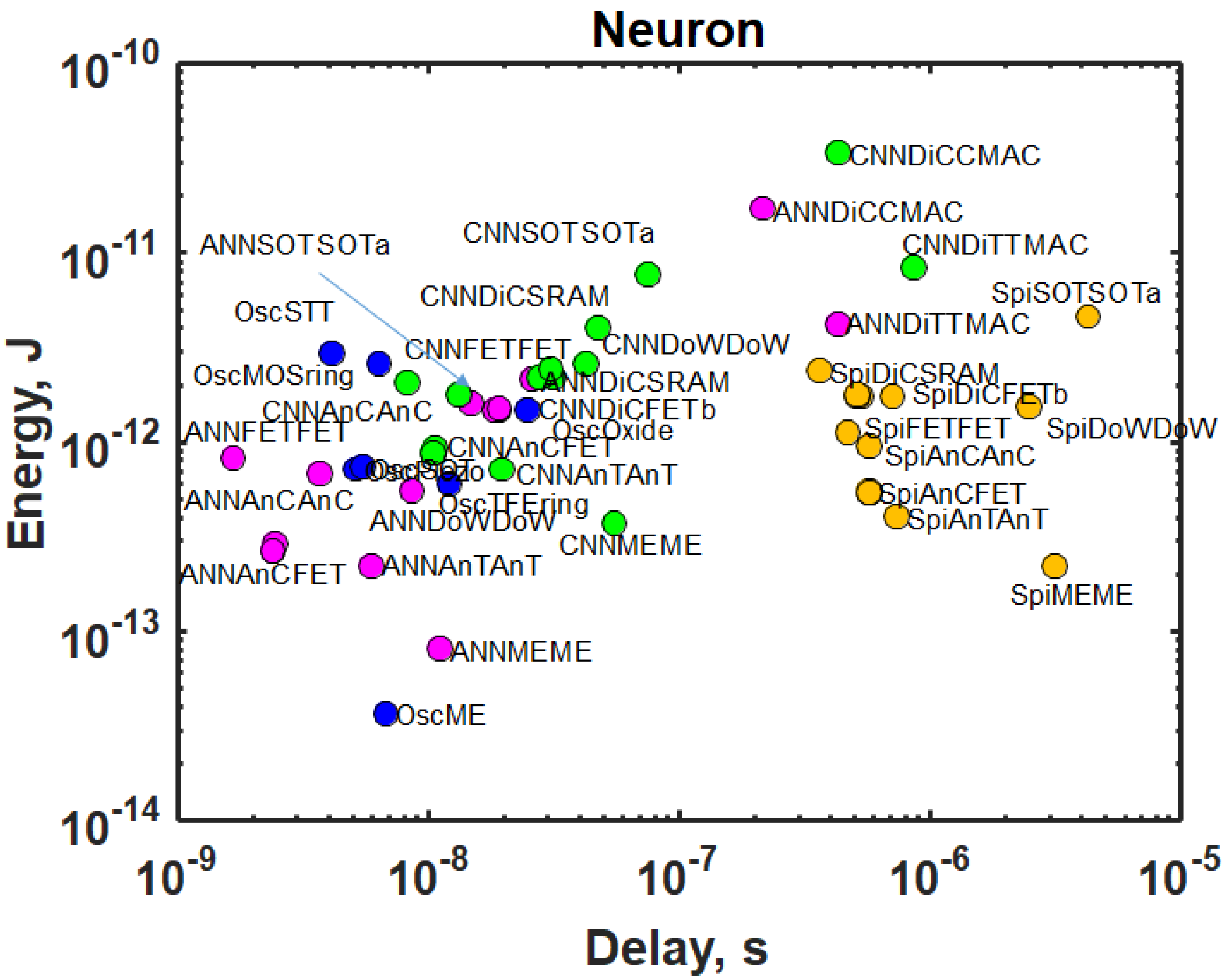

Figure 16. Energy vs. delay for a Neuron in ANN (magenta dots), CeNN (green dots), SNN (gold dots), ONN (blue dots). Labels for architectures according to Table 3.

The relative relation between energies and delays for a whole computing workload (LeNet in this case, Figure 17), closely resembles that for neurons. On the average, ANNs are faster than ONNs by about a half an  order of magnitude, faster than CNN by an order of magnitude, and faster than SNN by two orders of magnitude. Magnetoelectric devices are more energy efficient than analog neurons by about an order of magnitude. They are more energy efficient than digital neurons by about another order of magnitude. The redeeming quality of SNNs is built-in learning via spike-dependent timing plasticity (SDTP) [4].

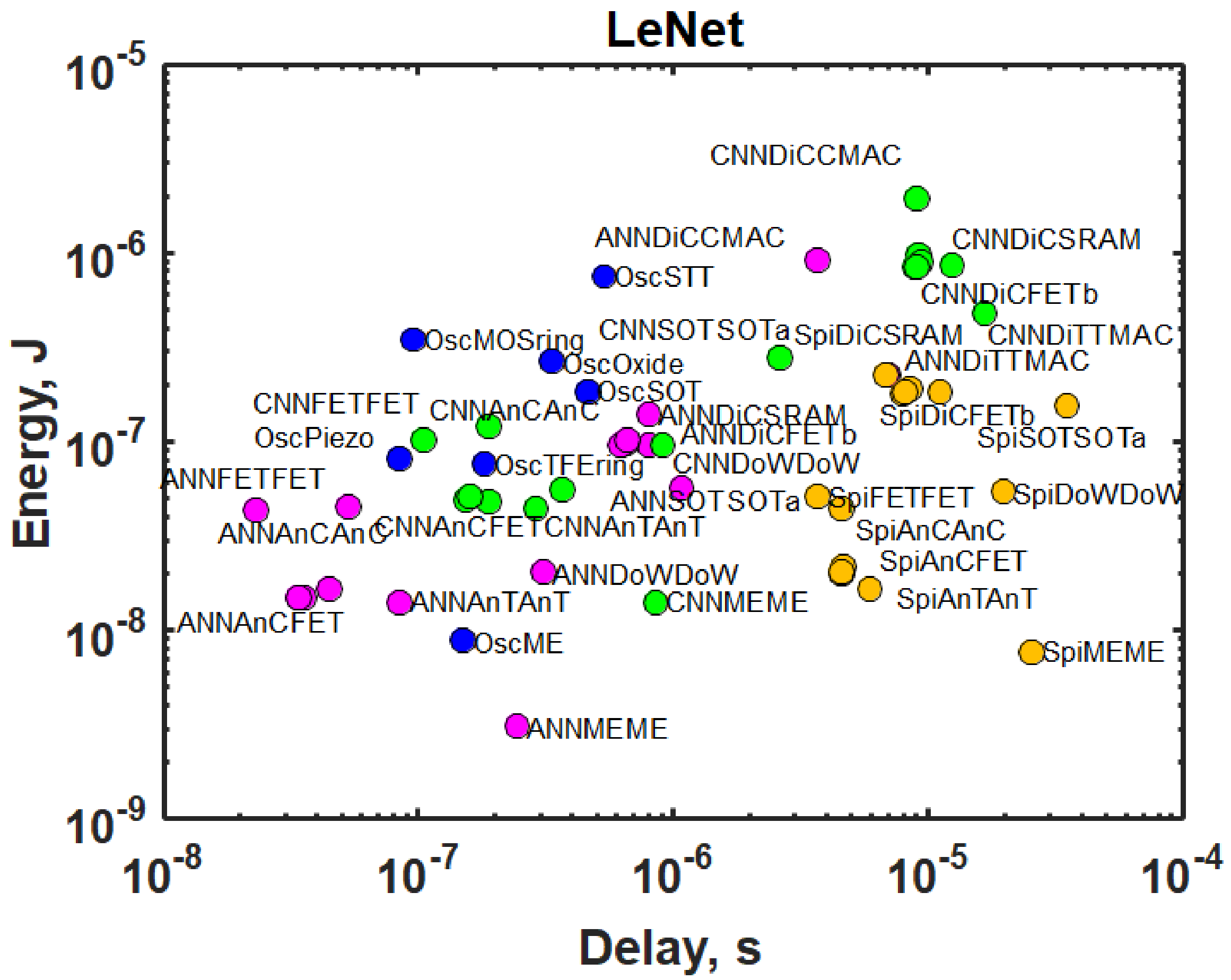

Figure 17. Energy vs. delay for one inference in a circuit implementing the LeNet convolutional neural network: ANN (magenta dots), CeNN (green dots), SNN (gold dots), ONN (blue dots). Labels for architectures according to Table 3. Bottoms-up benchmarks.

Now we include tops-down benchmarks for experimentally demonstrated integrated neuromorphic chips and neural accelerators, Figure 18. We notice that neural accelerators are within an order-of-magnitude of agreement with the bottoms-up benchmarks of their similar technology (MAC) implementation. Neuromorphic spiking chips in general prove to be slower than digital accelerators. The difference depends whether they are designed to run at biologically feasible firing rates (few tens of Hertz, e.g. ROLLS) or at an accelerated rate (tens of kilo-Hertz, e.g. HICANN). These are still slower than the clock rates of hundreds of mega-Hertz use in neural accelerators. Neuromorphic chips have a lower power of operation than neural accelerators. However their speed (determined by the firing rate) is much slower than that of neural accelerators (determined by the clock rate). As a result, the energy per inference (proportional to the product of the operation delay and power, proves to be higher in neuromorphic chips.

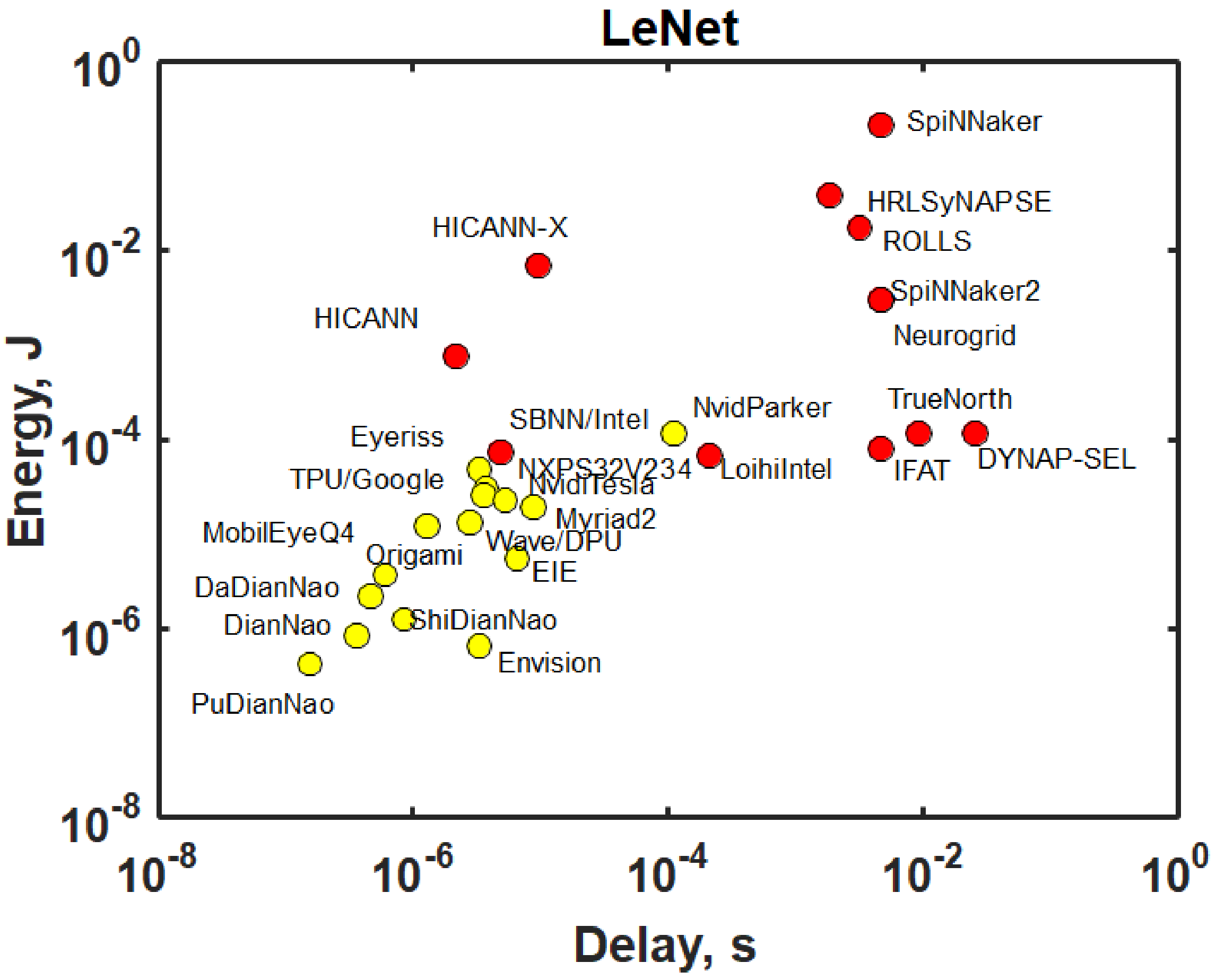

Figure 18. Energy vs. delay for one inference for CMOS circuit implementations of the LeNet convolutional neural network: with various digital accelerators (yellow dots) and neuromorphic spiking chips (red dots). Tops-down benchmarks. Data from Tables II and III are used.

Energy and delay for various workloads (but the same hardware) is shown in Figure 19. The numerical values are determined by the size of the overall network, mostly the number of MACs in it, see Supplementary materials. The relation between energy and delay between the various hardwares looks similar from workload to workload.

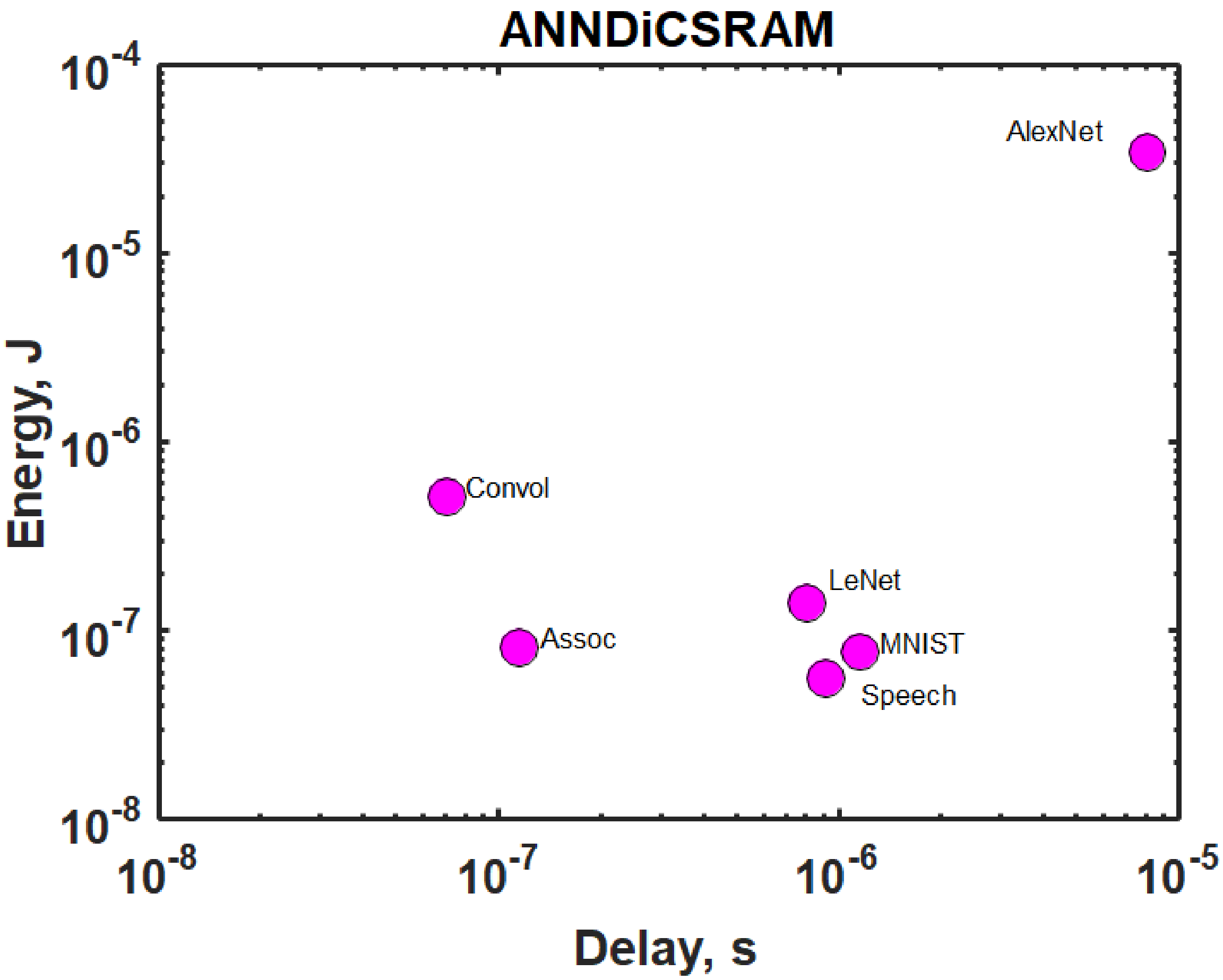

Figure 19. Energy vs. delay for one inference in various workloads implemented with digital neurons and SRAM synapses.

## 11.Throughput and Dissipated Power

Circuit performance can be represented as computing throughput plotted vs. dissipated power (Figure 20). One notices that spintronic networks have a higher per unit area throughput due to the small size of their implementation of neurons and synapses. However this higher throughput results in very high dissipated power. If we apply a cap on allowed power dissipation (Figure 21), some architectures with leading throughput values (e.g. ANNFETFET) are scaled proportionally. In this case only low-energy spintronic options maintain high throughput (e.g. ANNanCFET).

va

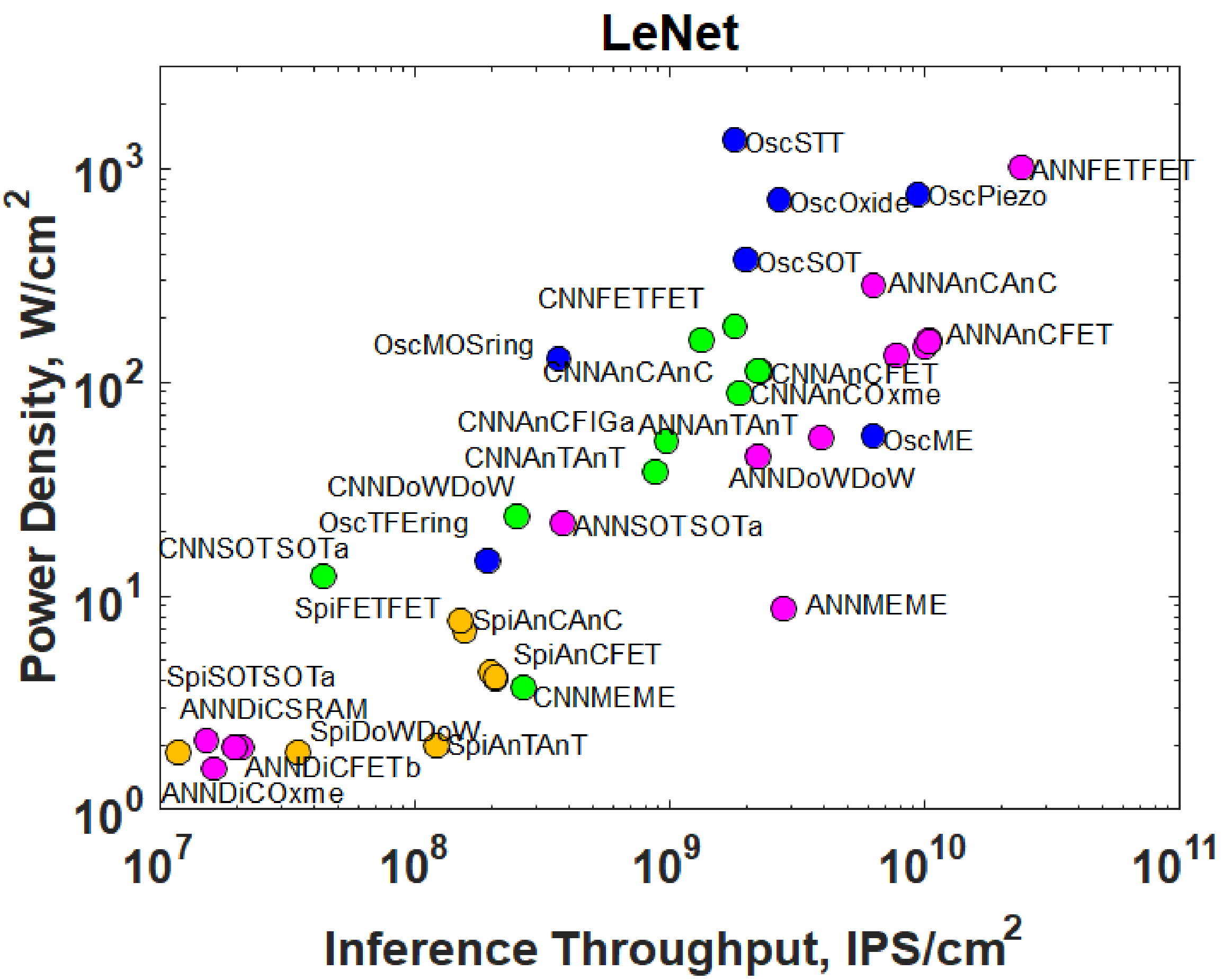

Figure 20. Dissipated power density vs. inference operation throughput per unit area in a circuit implementing the LeNet convolutional neural network. Labels for architectures according to Table 3.

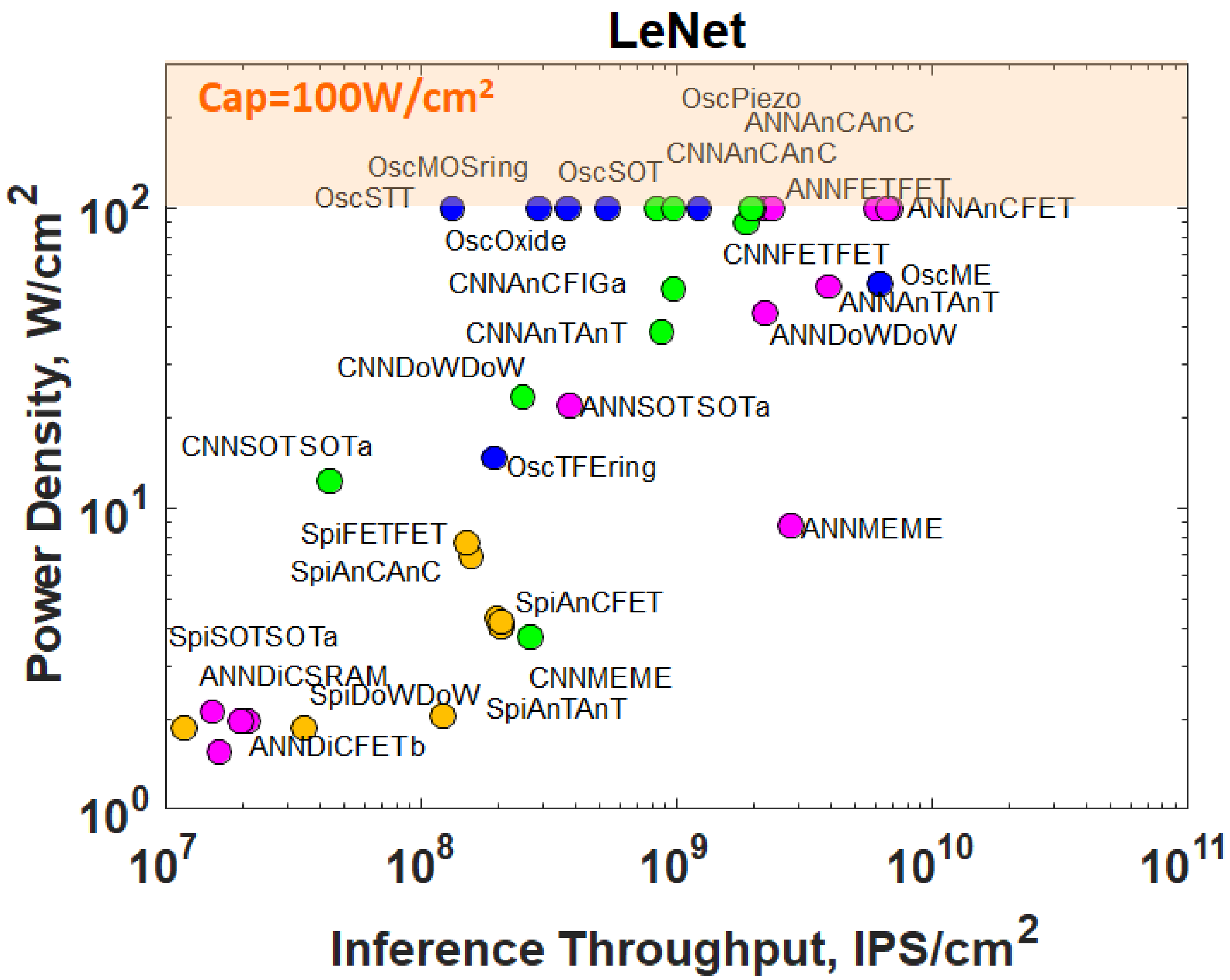

Figure 21. Dissipated power density vs. inference operation throughput per unit area in a circuit implementing the LeNet convolutional neural network. Power is capped to 100W/cm $^2$, throughput is scaled proportionally. Labels for architectures according to Table I.

## 12. Conclusions

In summary, the developed methodology described in this paper enables quantifying the effect of devices and NN types on the performance, power, and area of NNs. ANN and ONN show higher speed of operation at comparable energy vs. CeNN and SNN. This translates into a larger inference throughput especially under the limitation of power dissipation. The trend is confirmed by a comparison of actual fabricated functional neuromorphic chips (which are SNN) and neural accelerator chips (which are ANN based). Within each NN type, the ones based on multipliers and adders (MAC) prove to be less efficient while ones based on analog neurons and synapses prove to be more efficient in both speed and energy of operation. Among those, ferroelectric devices show higher speed and spintronic devices (especially based on magnetoelectric switching) show lower energy of operation.

It is important to note that the conclusions relate to inference and do not cover learning. Spiking neural networks are especially amenable to unsupervised learning; and this advantage is not comprehended in the present benchmarks.

## 13. Acknowledgements

The authors gratefully acknowledge discussions and critique by Narayan Srinivasa, Mike Mayberry, Sasikanth Manipatruni, Greg Chen, Ram Krishnamurthy, Chenyun Pan, Azad Naeemi, Dan Hammerstrom, Mike Davies, Eugenio Culurciello, Dmitri Strukov, Kaushik Roy, and Wolfgang Porod.

## 14. Supplementary Materials

Remaining benchmarking plots are collected here in order to keep the main text concise.

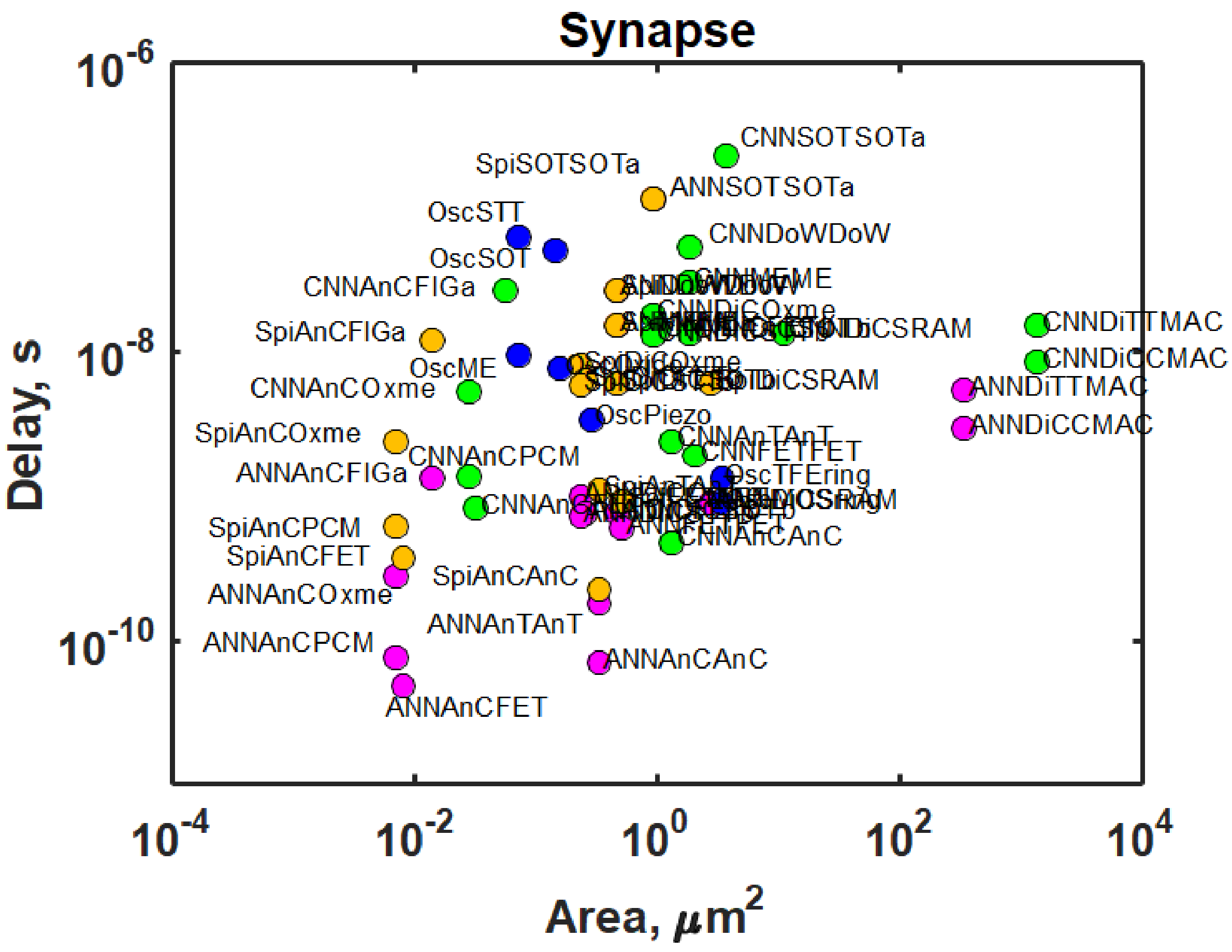

Figure 22. Delay vs. area for synapses.

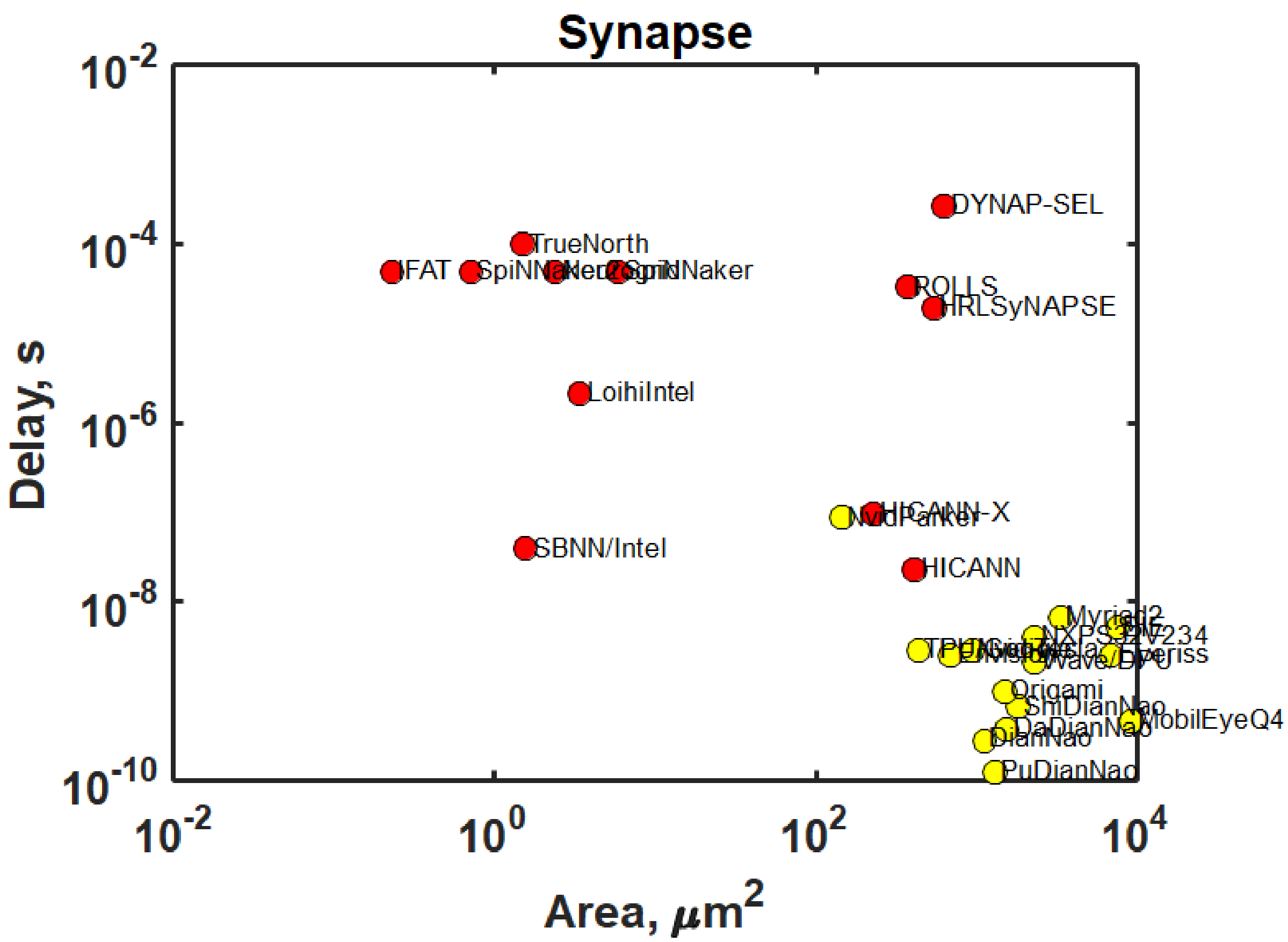

Figure 23. Delay vs. area for synapses.

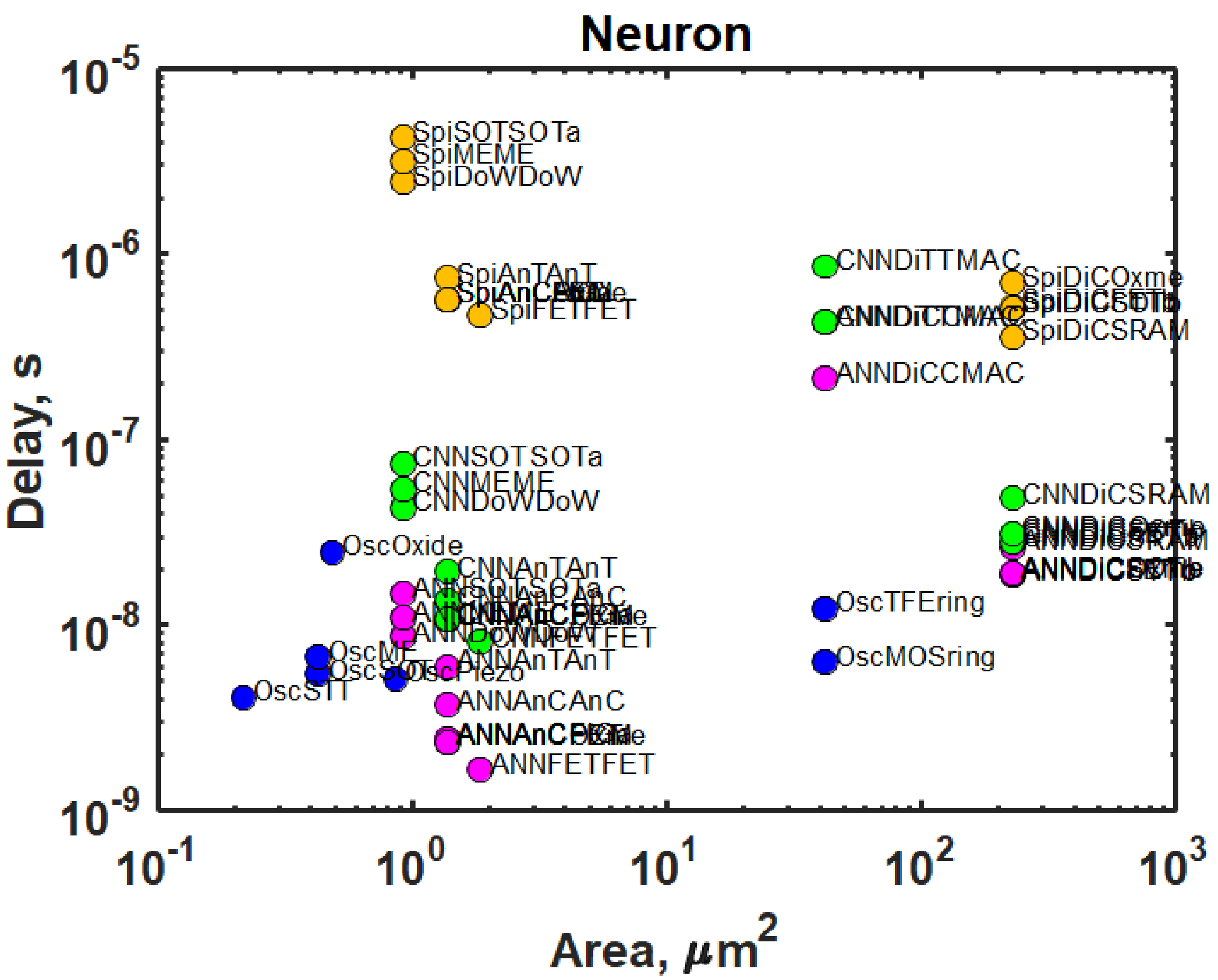

Figure 24. Delay vs. area for neurons.

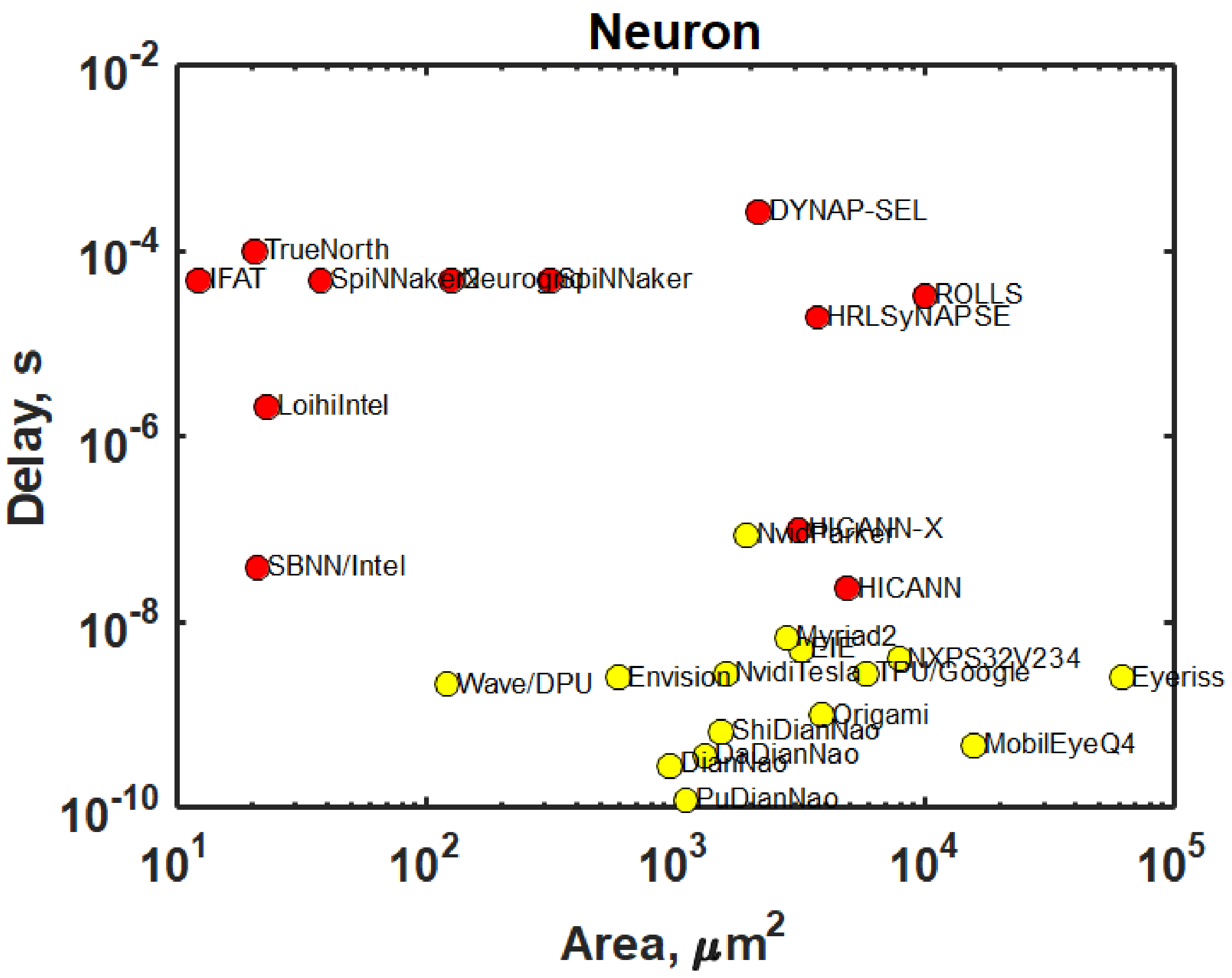

Figure 25. Delay vs. area for neurons.

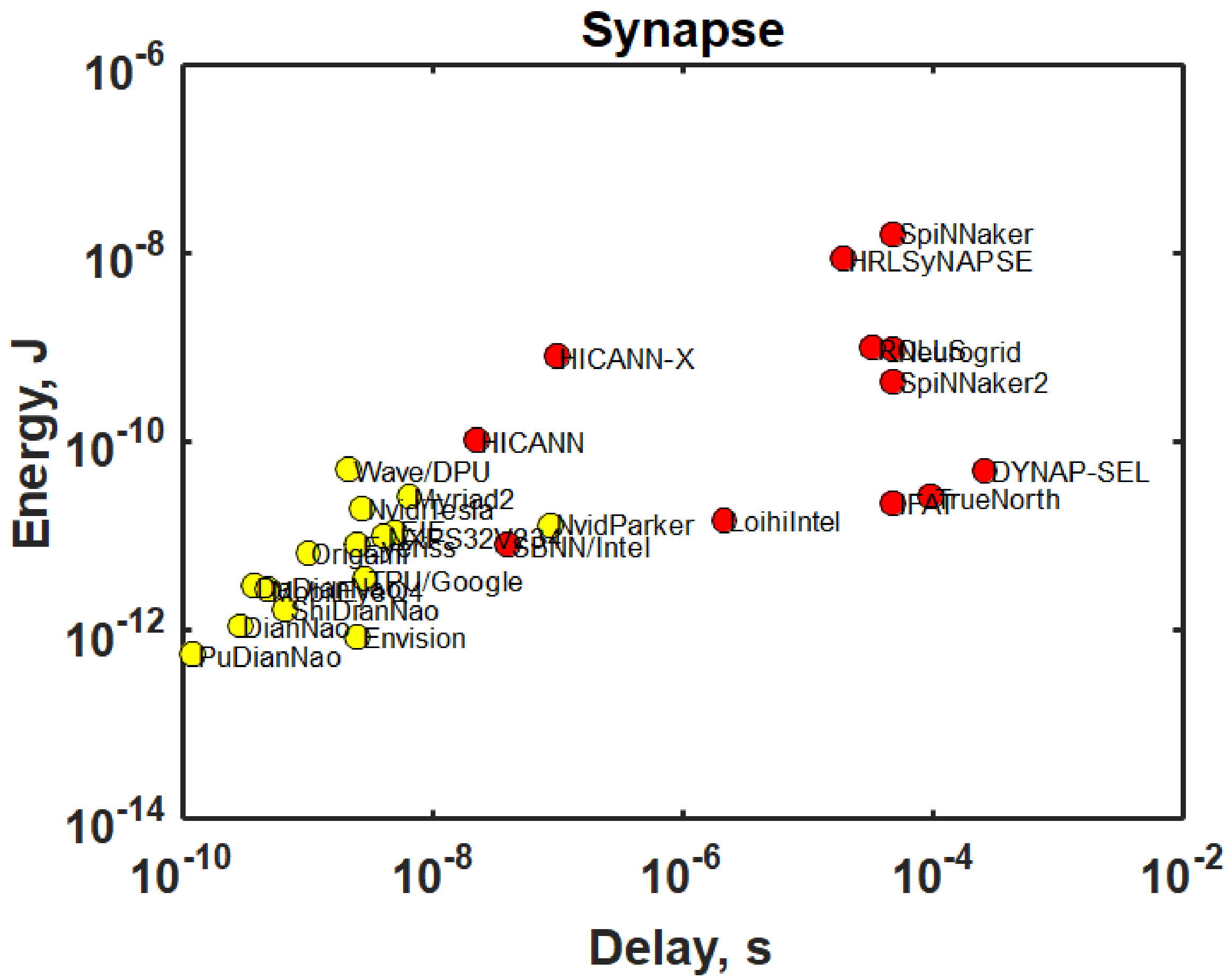

Figure 26. Energy vs. delay for synapses.

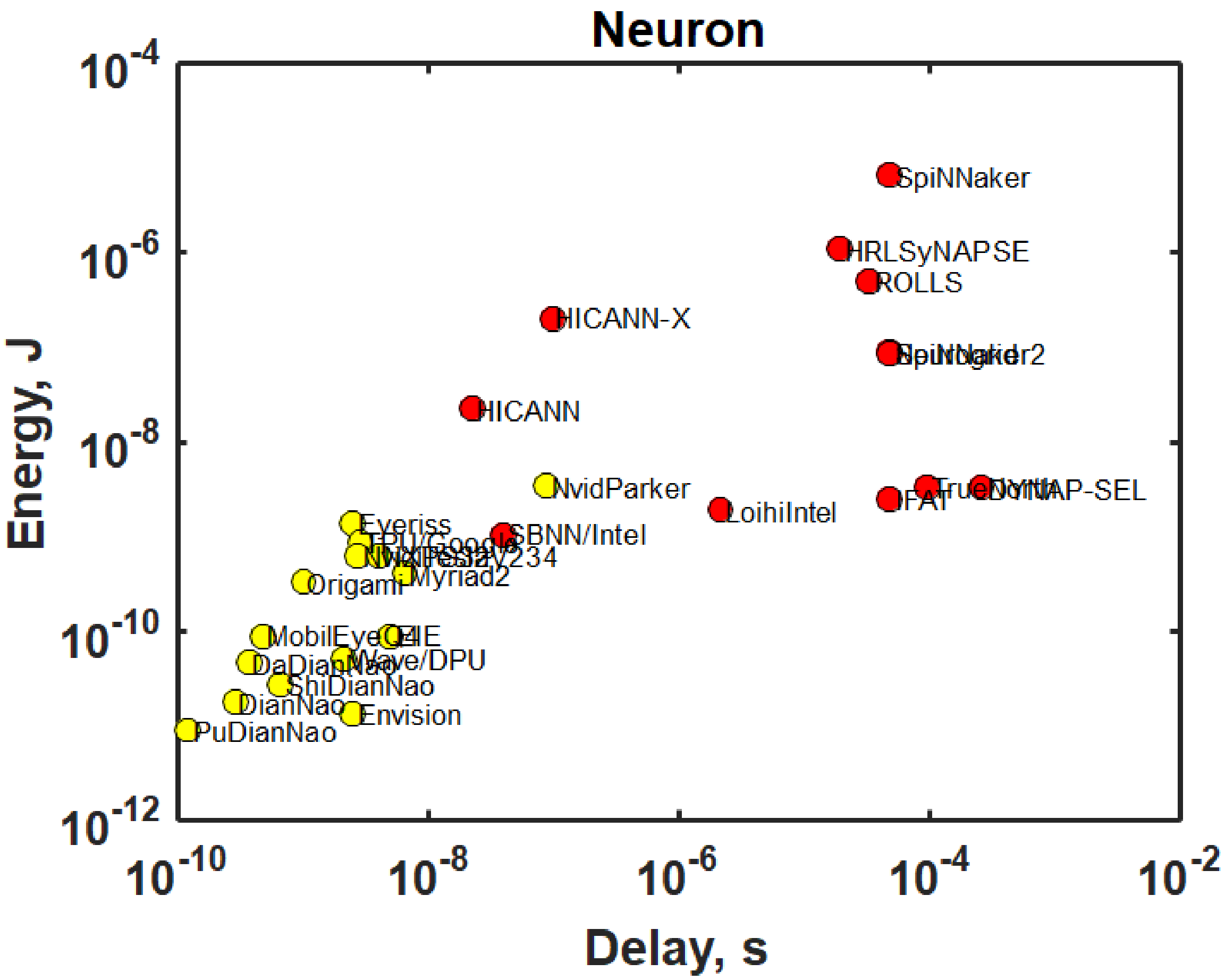

Figure 27. Energy vs. delay for neurons.

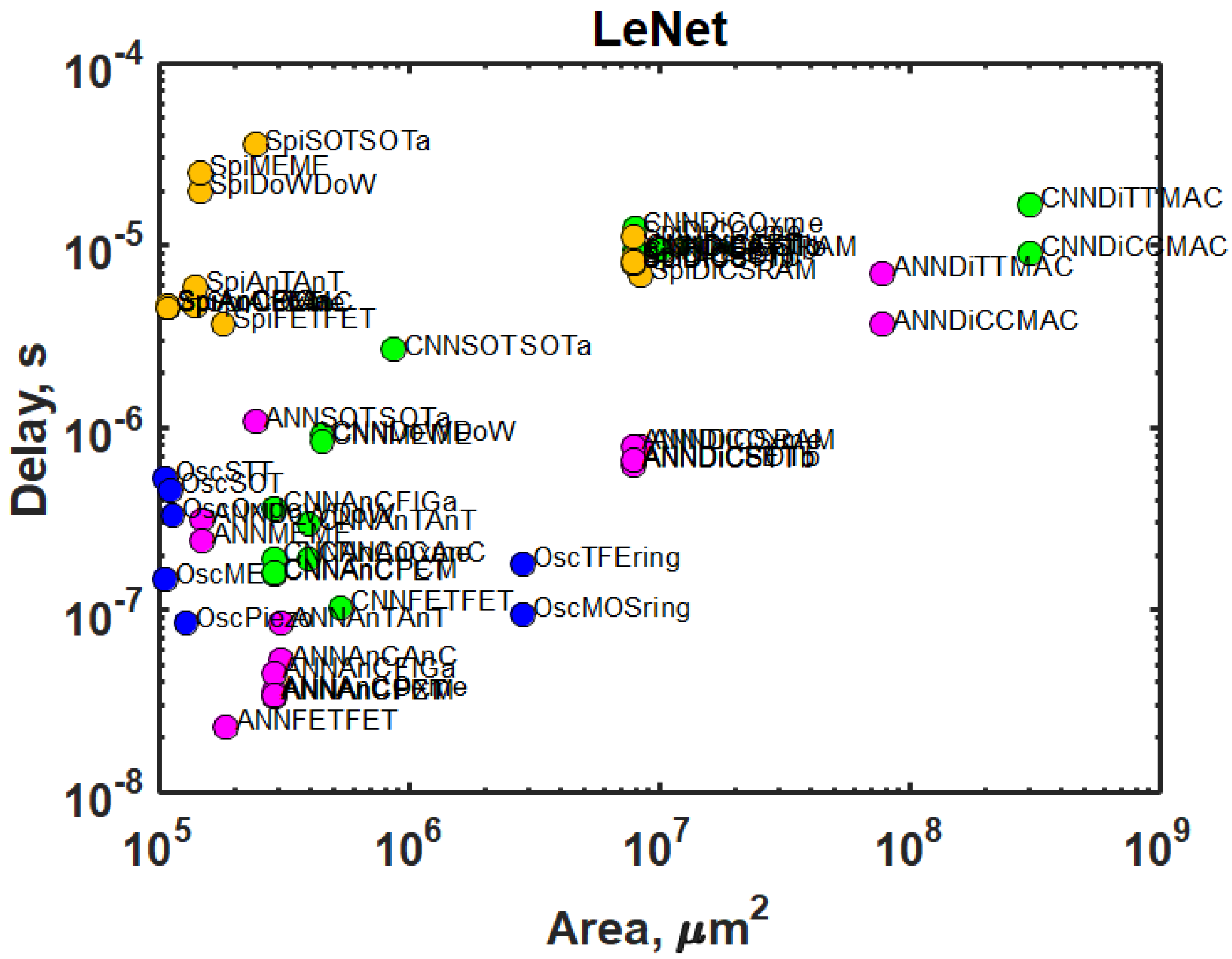

Figure 28. Delay vs. area for LeNet CoNN.

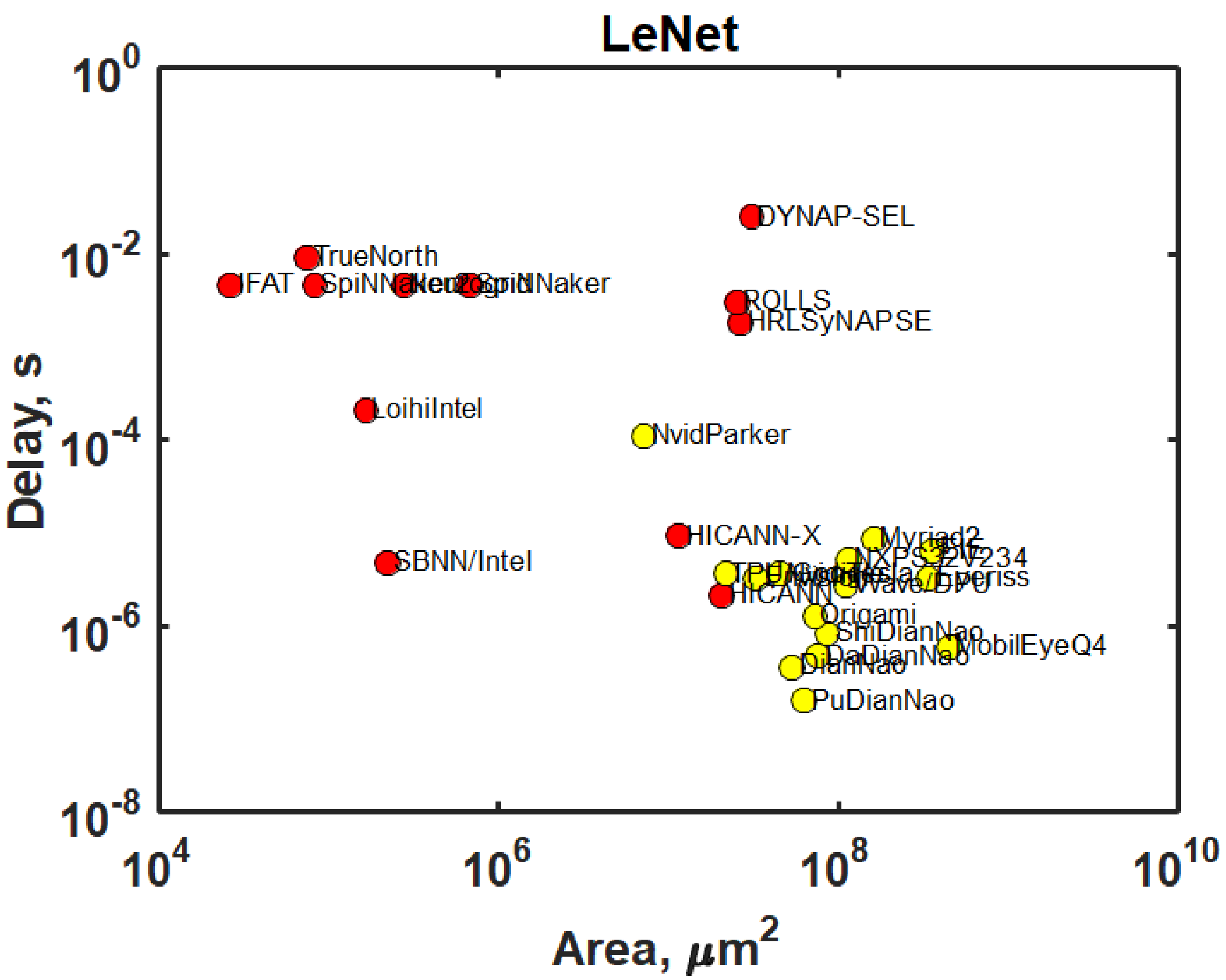

Figure 29. Delay vs. area for LeNet CoNN.

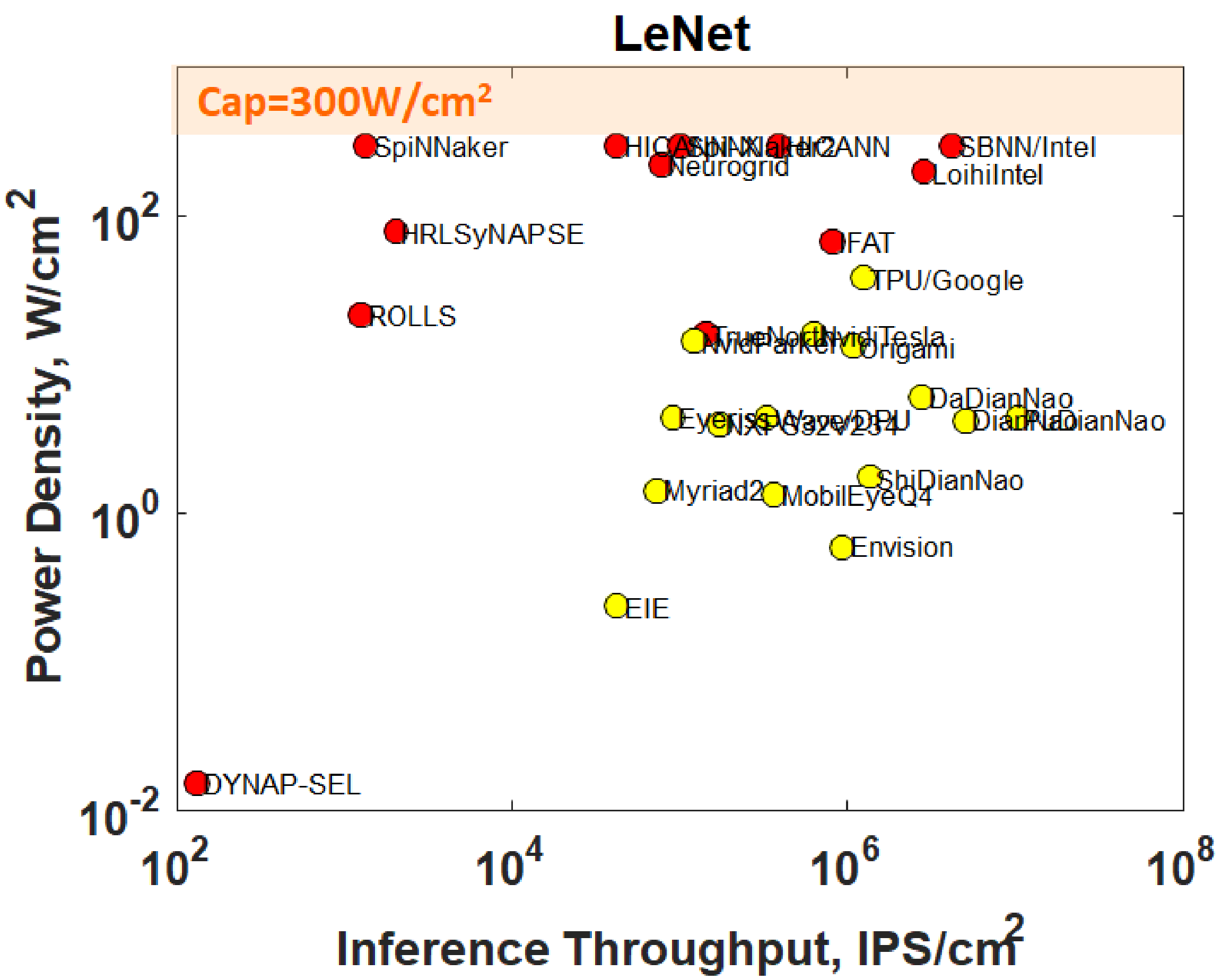

Figure 30. Dissipated power density vs. inference operation throughput per unit area in a circuit implementing the LeNet convolutional neural network, includes benchmarks for prototype neuromorphic chips and neural accelerators.

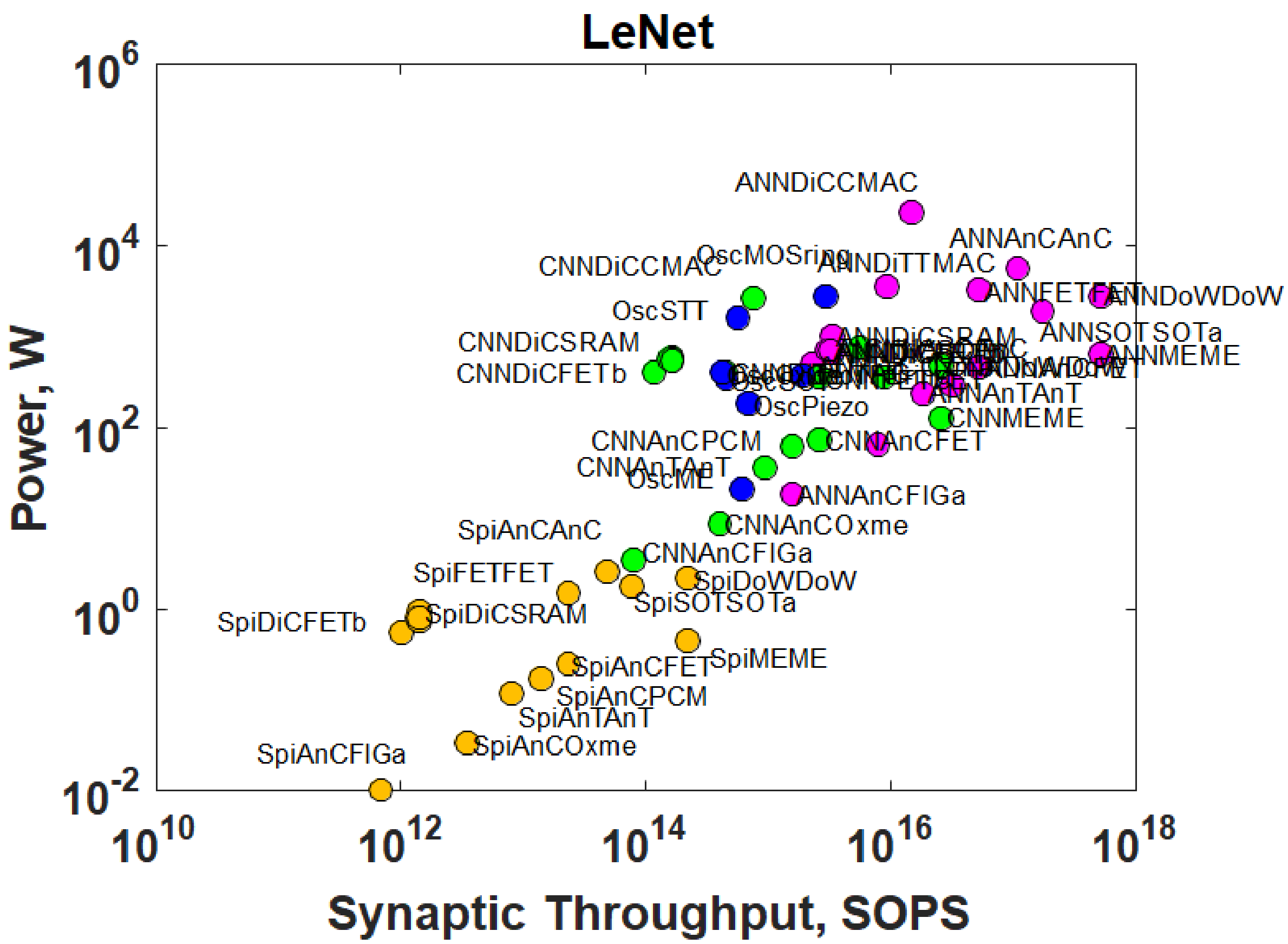

Figure 31. Power vs. synaptic throughput for LeNet.

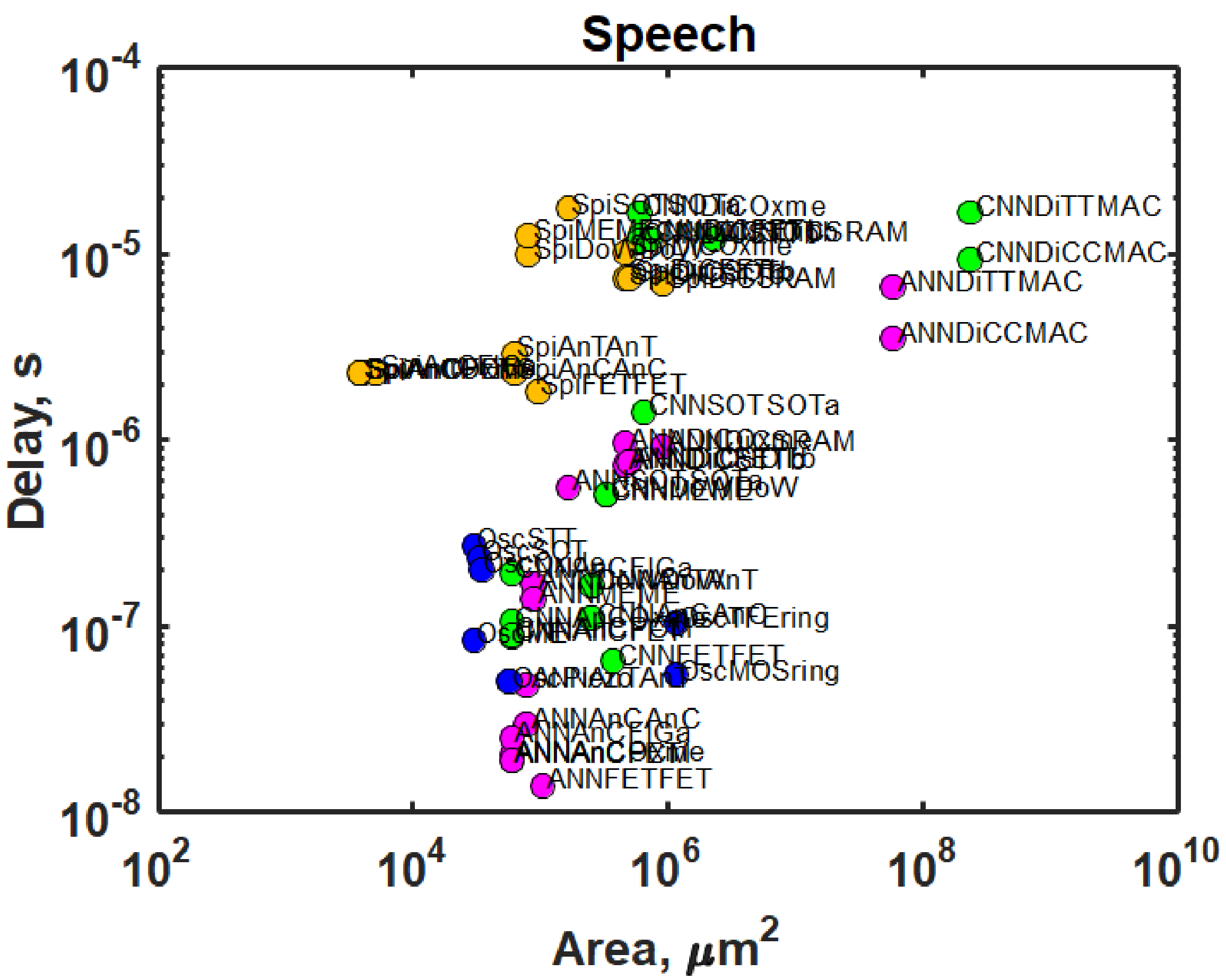

Figure 32. Delay vs. area for the speech recognition.

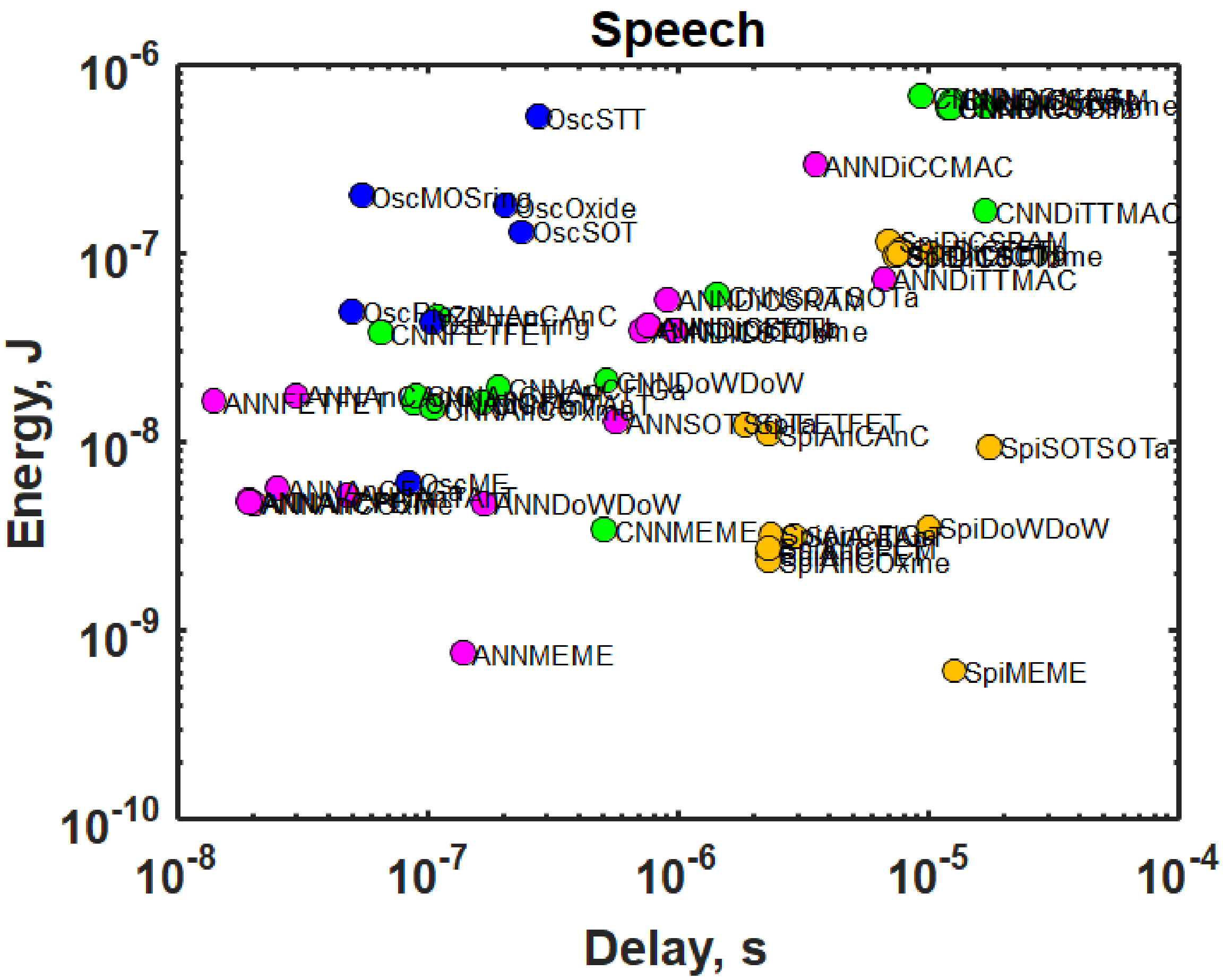

Figure 33. Energy vs. delay for the speech recognition.

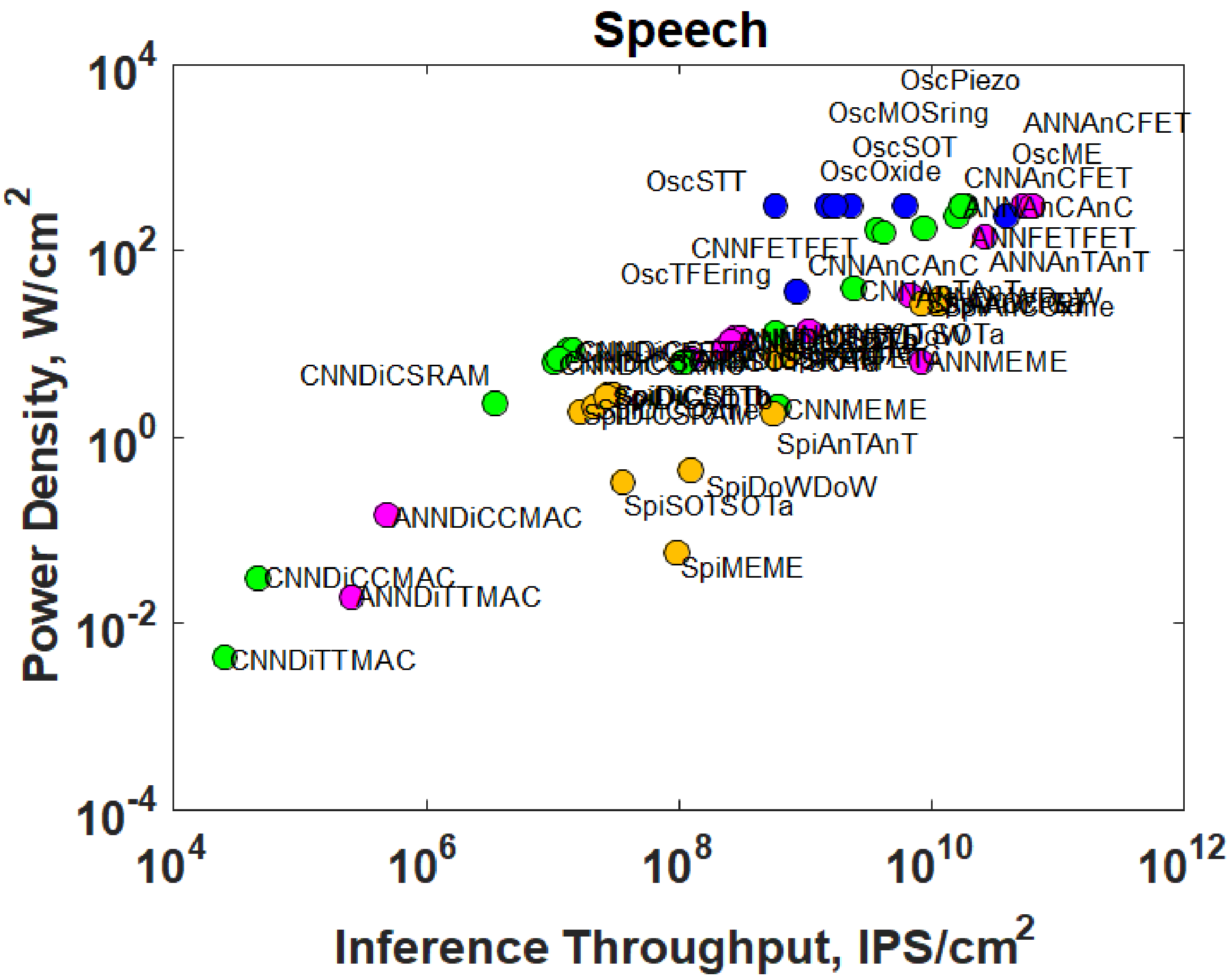

Figure 34. Power density vs. inference throughput for the speech recognition.

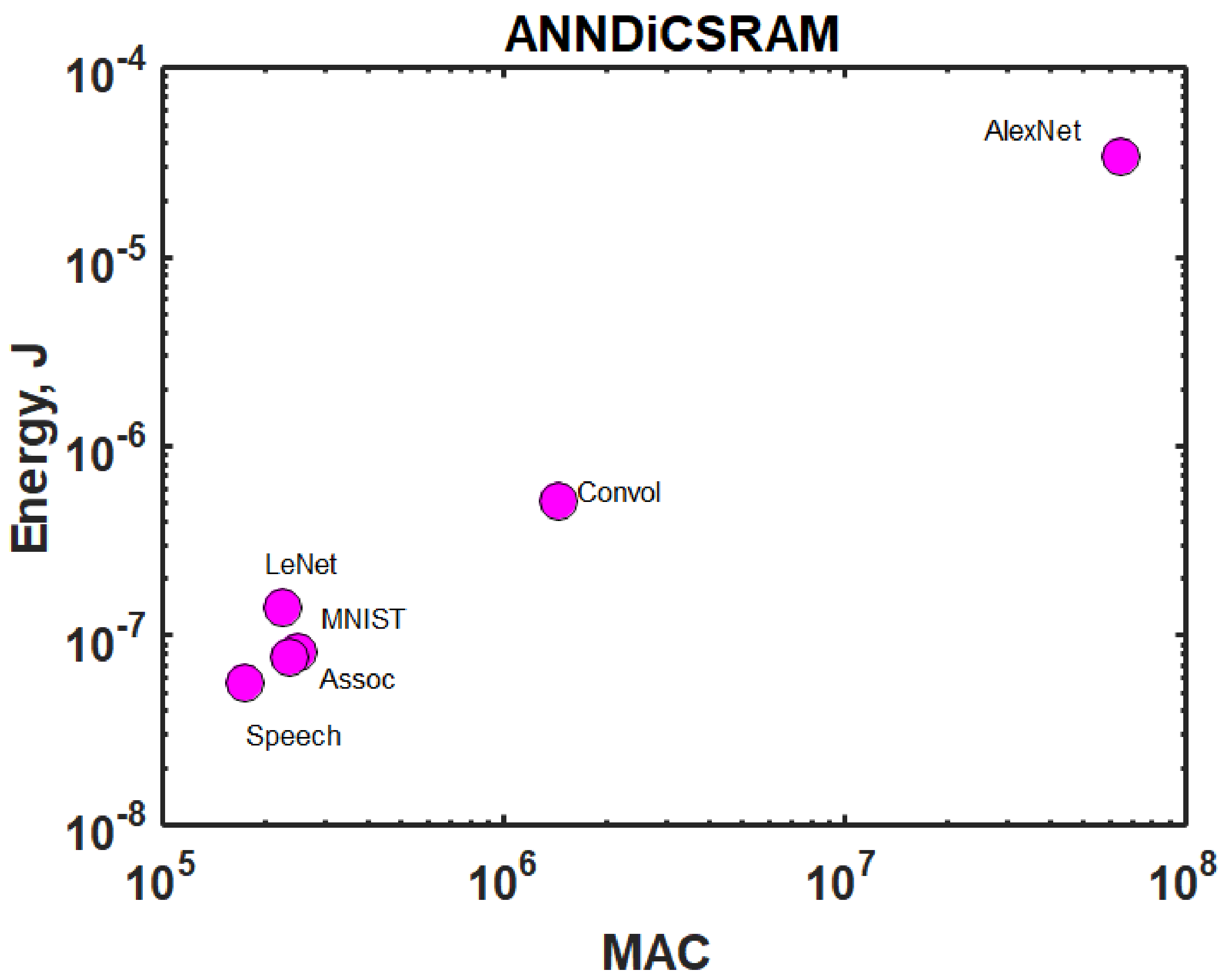

Figure 35. Energy vs. MAC in digital neurons and SRAM synapses for various workloads.

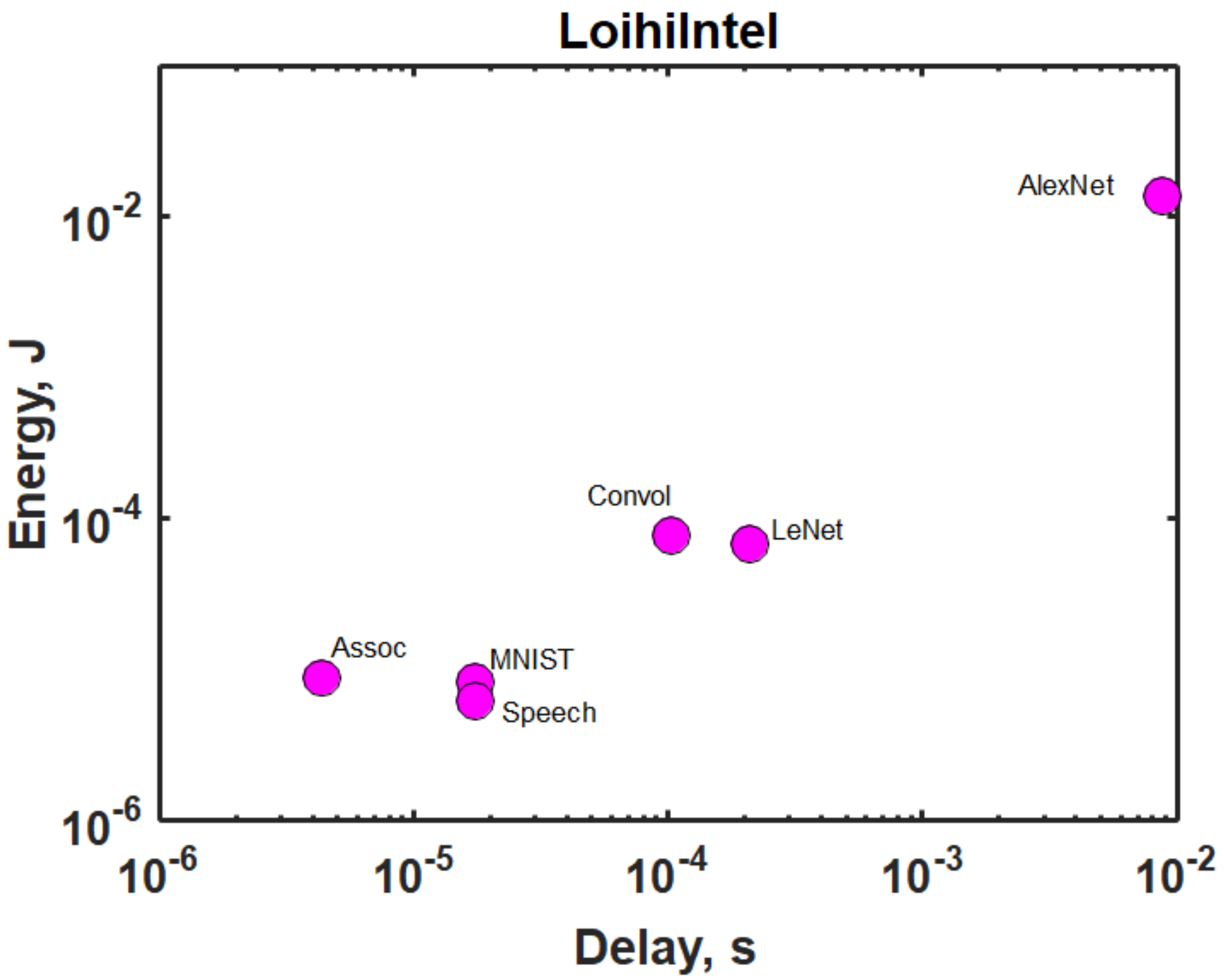

Figure 36. Energy vs. delay in Loihi for various workloads.

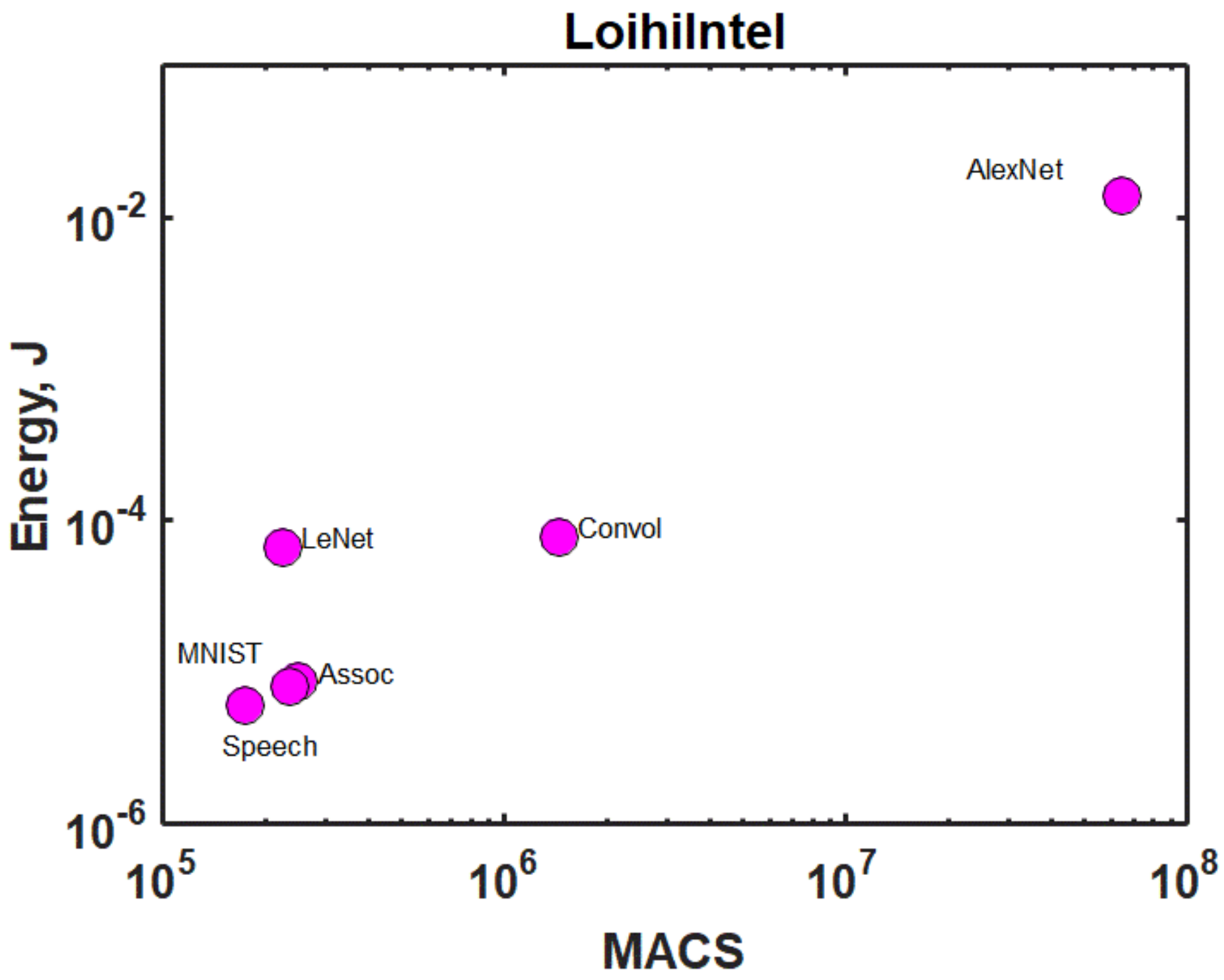

Figure 37. Energy vs. MAC in Loihi for various workloads.

Table 7. Performance benchmarks for the combinations of devices and network types.

| hardware | Area, syn | Area, lic | Area, neu | Area, gic | Delay, syn | Delay, lic | Delay, neu | Delay, gic | Energy, syn | Energy, lic | Energy, neu | Energy, gic |
|---|---|---|---|---|---|---|---|---|---|---|---|---|
| units | $um^2$ | $um^2$ | $um^2$ | $um^2$ | ps | ps | ps | ps | aJ | aJ | aJ | aJ |
| ANNDiCSRAM | 2.765 | 25.540 | 228.290 | 370.610 | 897.31 | 252.79 | 26144.00 | 24981.00 | 306.81 | 136.21 | 2115.30 | 1976.60 |
| ANNDiCCMAC | 336.900 | 281.930 | 42.034 | 3191.200 | 2932.60 | 2790.50 | 215260.00 | 215100.00 | 1531.10 | 1503.60 | 17034.00 | 17020.00 |
| ANNDiCOxme | 0.230 | 7.373 | 229.850 | 247.360 | 980.62 | 72.97 | 19083.00 | 16673.00 | 211.65 | 39.32 | 1457.90 | 1319.20 |
| ANNDiTTMAC | 336.900 | 281.930 | 42.034 | 3191.200 | 5378.90 | 5148.00 | 425740.00 | 425490.00 | 381.56 | 375.90 | 4257.90 | 4254.90 |
| ANNDiCFETb | 0.259 | 7.820 | 229.850 | 249.110 | 759.04 | 77.40 | 18572.00 | 16791.00 | 221.87 | 41.71 | 1467.20 | 1328.60 |
| ANNDiCSTTb | 0.230 | 7.373 | 229.850 | 247.360 | 718.72 | 72.97 | 18385.00 | 16673.00 | 206.42 | 39.32 | 1457.90 | 1319.20 |
| ANNDiCSOTb | 0.461 | 10.427 | 229.850 | 261.040 | 756.92 | 103.20 | 19327.00 | 17596.00 | 221.08 | 55.61 | 1530.90 | 1392.20 |
| ANNAnCAnC | 0.338 | 8.923 | 1.382 | 102.560 | 70.46 | 51.32 | 3716.90 | 1728.30 | 49.54 | 47.59 | 685.28 | 546.98 |
| ANNAnTAnT | 0.338 | 8.923 | 1.382 | 102.560 | 178.37 | 64.65 | 5951.30 | 3418.60 | 12.57 | 11.90 | 225.78 | 136.75 |
| ANNAnCFET | 0.008 | 1.382 | 1.382 | 23.891 | 47.96 | 7.95 | 2391.30 | 402.59 | 7.89 | 7.37 | 265.72 | 127.42 |
| ANNAnCOxme | 0.007 | 1.303 | 1.382 | 23.315 | 273.51 | 7.50 | 2381.60 | 392.89 | 7.21 | 6.95 | 262.65 | 124.35 |
| ANNAnCFlGa | 0.014 | 1.843 | 1.382 | 27.586 | 1340.20 | 10.60 | 2453.60 | 464.87 | 10.88 | 9.83 | 285.43 | 147.13 |
| ANNAnCPCM | 0.007 | 1.303 | 1.382 | 23.315 | 74.10 | 7.50 | 2381.60 | 392.89 | 8.00 | 6.95 | 262.65 | 124.35 |
| ANNFETFET | 0.518 | 11.059 | 1.843 | 126.850 | 595.99 | 555.97 | 1657.00 | 44.41 | 59.50 | 58.98 | 825.00 | 676.51 |
| ANNDoWDoW | 0.461 | 10.427 | 0.922 | 118.880 | 27027.00 | 27023.00 | 8586.20 | 40.95 | 3.57 | 3.48 | 550.80 | 39.63 |
| ANNSOTSOTa | 0.922 | 14.746 | 0.922 | 167.480 | 114340.00 | 114330.00 | 14889.00 | 33.23 | 4.96 | 4.92 | 1586.60 | 55.83 |
| ANNMEME | 0.461 | 10.427 | 0.922 | 118.880 | 15090.00 | 15086.00 | 11060.00 | 129.91 | 0.96 | 0.87 | 80.88 | 9.91 |
| CNNDiCSRAM | 11.06 | 51.08 | 228.29 | 622.77 | 13396.00 | 505.58 | 47791.00 | 41978.00 | 3684.40 | 272.43 | 4014.70 | 3321.40 |
| CNNDiCCMAC | 1347.60 | 563.86 | 42.03 | 6380.10 | 8422.80 | 5580.90 | 430820.00 | 430050.00 | 3557.80 | 3007.20 | 34098.00 | 34027.00 |
| CNNDiCOxme | 0.92 | 14.75 | 229.85 | 286.46 | 18299.00 | 145.95 | 31358.00 | 19309.00 | 3525.10 | 78.64 | 2221.10 | 1527.80 |
| CNNDiTTMAC | 1347.60 | 563.86 | 42.03 | 6380.10 | 14914.00 | 10296.00 | 851930.00 | 850680.00 | 864.97 | 751.81 | 8521.40 | 8506.80 |
| CNNDiCFETb | 1.04 | 15.64 | 229.85 | 292.47 | 13788.00 | 154.80 | 28616.00 | 19714.00 | 3686.60 | 83.41 | 2253.10 | 1559.80 |
| CNNDiCSTTb | 0.92 | 14.75 | 229.85 | 286.46 | 13061.00 | 145.95 | 27868.00 | 19309.00 | 3420.70 | 78.64 | 2221.10 | 1527.80 |
| CNNDiCSOTb | 1.84 | 20.85 | 229.85 | 331.50 | 13281.00 | 206.40 | 31002.00 | 22345.00 | 3420.60 | 111.22 | 2461.30 | 1768.00 |
| CNNAnCAnC | 1.35 | 17.85 | 1.38 | 202.72 | 485.40 | 102.64 | 13360.00 | 3416.10 | 134.13 | 95.18 | 1772.70 | 1081.20 |
| CNNAnTAnT | 1.35 | 17.85 | 1.38 | 202.72 | 2403.70 | 129.29 | 19421.00 | 6757.30 | 37.16 | 23.80 | 715.46 | 270.29 |
| CNNAnCFET | 0.03 | 2.76 | 1.38 | 36.12 | 816.14 | 15.90 | 10552.00 | 608.66 | 25.01 | 14.75 | 884.14 | 192.64 |
| CNNAnCOxme | 0.03 | 2.61 | 1.38 | 34.58 | 5335.30 | 14.99 | 10526.00 | 582.74 | 19.14 | 13.90 | 875.94 | 184.43 |
| CNNAnCFlGa | 0.06 | 3.69 | 1.38 | 45.45 | 26613.00 | 21.20 | 10709.00 | 765.88 | 40.60 | 19.66 | 933.90 | 242.39 |
| CNNAnCPCM | 0.03 | 2.61 | 1.38 | 34.58 | 1347.00 | 14.99 | 10526.00 | 582.74 | 34.84 | 13.90 | 875.94 | 184.43 |
| CNNFETFET | 2.07 | 22.12 | 1.84 | 251.11 | 1912.20 | 1111.90 | 8150.70 | 87.92 | 128.22 | 117.96 | 2081.70 | 1339.20 |
| CNNDoWDoW | 1.84 | 20.85 | 0.92 | 236.39 | 54127.00 | 54045.00 | 42808.00 | 81.43 | 8.84 | 6.95 | 2634.70 | 78.80 |
| CNNSOTSOTa | 3.69 | 29.49 | 0.92 | 333.98 | 228900.00 | 228650.00 | 74347.00 | 66.26 | 10.67 | 9.83 | 7765.20 | 111.33 |
| CNNMEME | 1.84 | 20.85 | 0.92 | 236.39 | 30254.00 | 30172.00 | 54911.00 | 258.32 | 3.62 | 1.74 | 374.54 | 19.70 |
| SpiDiCSRAM | 2.76 | 25.54 | 228.29 | 370.61 | 6053.40 | 252.79 | 359810.00 | 24981.00 | 648.01 | 136.21 | 2392.60 | 1976.60 |
| SpiDiCOxme | 0.23 | 7.37 | 229.85 | 247.36 | 8241.70 | 72.97 | 710690.00 | 16673.00 | 556.30 | 39.32 | 1735.20 | 1319.20 |
| SpiDiCFETb | 0.26 | 7.82 | 229.85 | 249.11 | 6212.10 | 77.40 | 529560.00 | 16791.00 | 582.19 | 41.71 | 1744.50 | 1328.60 |
| SpiDiCSTTb | 0.23 | 7.37 | 229.85 | 247.36 | 5884.70 | 72.97 | 509660.00 | 16673.00 | 540.63 | 39.32 | 1735.20 | 1319.20 |
| SpiDiCSOTb | 0.46 | 10.43 | 229.85 | 261.04 | 5986.70 | 103.20 | 516240.00 | 17596.00 | 552.02 | 55.61 | 1808.20 | 1392.20 |
| SpiAnCAnC | 0.34 | 8.92 | 1.38 | 102.56 | 223.56 | 51.32 | 574470.00 | 1728.30 | 53.43 | 47.59 | 961.88 | 546.98 |
| SpiAnTAnT | 0.34 | 8.92 | 1.38 | 102.56 | 1088.10 | 64.65 | 732830.00 | 3418.60 | 13.90 | 11.90 | 403.85 | 136.75 |
| SpiAnCFET | 0.01 | 1.38 | 1.38 | 23.89 | 368.06 | 7.95 | 573150.00 | 402.59 | 8.91 | 7.37 | 542.32 | 127.42 |
| SpiAnCOxme | 0.01 | 1.30 | 1.38 | 23.32 | 2401.60 | 7.50 | 573140.00 | 392.89 | 7.74 | 6.95 | 539.25 | 124.35 |
| SpiAnCFlGa | 0.01 | 1.84 | 1.38 | 27.59 | 11977.00 | 10.60 | 573210.00 | 464.87 | 12.97 | 9.83 | 562.03 | 147.13 |
| SpiAnCPCM | 0.01 | 1.30 | 1.38 | 23.32 | 606.90 | 7.50 | 573140.00 | 392.89 | 10.09 | 6.95 | 539.25 | 124.35 |
| SpiFETFET | 0.52 | 11.06 | 1.84 | 126.85 | 916.08 | 555.97 | 464460.00 | 44.41 | 60.52 | 58.98 | 1122.00 | 676.51 |
| SpiDoWDoW | 0.46 | 10.43 | 0.92 | 118.88 | 27060.00 | 27023.00 | 2461100.00 | 40.95 | 3.76 | 3.48 | 1573.20 | 39.63 |
| SpiSOTSOTa | 0.92 | 14.75 | 0.92 | 167.48 | 114440.00 | 114330.00 | 4278600.00 | 33.23 | 5.04 | 4.92 | 4648.10 | 55.83 |
| SpiMEME | 0.46 | 10.43 | 0.92 | 118.88 | 15123.00 | 15086.00 | 3148100.00 | 129.91 | 1.15 | 0.87 | 222.81 | 9.91 |
| OscMOSring | 3.38 | 28.22 | 41.47 | 334.22 | 876.11 | 162.28 | 6346.00 | 5632.10 | 974.47 | 150.50 | 2606.50 | 1782.50 |
| OscTFEring | 3.38 | 28.22 | 41.47 | 334.22 | 1320.00 | 204.43 | 12256.00 | 11141.00 | 204.01 | 37.62 | 612.02 | 445.63 |
| OscPiezo | 0.29 | 8.24 | 0.86 | 94.35 | 3434.50 | 414.40 | 5134.10 | 2114.00 | 252.74 | 43.96 | 711.96 | 503.18 |
| OscSTTfast | 0.07 | 4.12 | 0.22 | 47.17 | 61174.00 | 57358.00 | 4095.90 | 279.51 | 2901.50 | 3.09 | 2933.80 | 35.38 |
| OscSOT | 0.14 | 5.83 | 0.43 | 66.71 | 49747.00 | 45192.00 | 5402.40 | 847.05 | 719.49 | 1.94 | 739.78 | 22.24 |
| OscME | 0.07 | 4.12 | 0.43 | 47.71 | 9362.80 | 5963.20 | 6736.20 | 3336.70 | 33.61 | 0.34 | 37.24 | 3.98 |
| OscOxide | 0.16 | 6.18 | 0.49 | 70.76 | 7628.60 | 2628.60 | 24655.00 | 19655.00 | 951.52 | 51.52 | 1489.70 | 589.66 |